\documentclass[12pt]{article}

\usepackage{psfig}

\newlength{\figwidth}
\newcommand{\eabe} {\begin{eqnarray}}
\newcommand{\eaen} {\end{eqnarray}}
\newcommand{\eqbe} {\begin{equation}}
\newcommand{\eqen} {\end{equation}}

\newcommand{\mrm} {\mathrm}
\newcommand{\srm}[1] {_{\mathrm{#1}}}
\newcommand{\ol} {\overline}
\renewcommand{\ln} {\mrm {ln}}
\newcommand{\al} {\alpha_0}
\newcommand{\alq} {\alpha\srm q}

\newcommand{\CF} {C\srm F}
\newcommand{\Nf} {N\srm f}
\newcommand{\Nc} {N\srm c}

\newcommand{\ra} {\rightarrow}
\newcommand{\dL} {\frac{\mrm d}{\mrm dL}}
\newcommand{\dLL} {\frac{\mrm d^2}{\mrm dL^2}}
\newcommand{\cP} {{\cal P}}

\newcommand{\nbar}[1] {{\ol {n^{(\mrm{#1})}}}}
\newcommand{\crqg}[1] {{c\srm r^{(\mrm{#1})}}}
\newcommand{\Nqq} {{N\srm{qq}^h}}
\newcommand{\Ngg} {{N\srm{gg}^h}}
\newcommand{\dcut} {{d\srm{cut}}}

\newcommand{\bibl}[5]
	{#1, {\it #2} {\bf #3} (#4) #5}
\newcommand{\anti}[1] {${ \ol \mrm #1 }$}
\newcommand{\pair}[1] {${\mrm {#1 \ol #1} }$}

\begin{document}

\begin{titlepage}
\begin{flushright}
 LU TP 98-11\\
 September 1998
\end{flushright}
\vspace{25mm}
\begin{center}
  \Large
  {\bf Energy and Virtuality Scale Dependence in Quark and Gluon Jets} \\
  \vspace{12mm}
  \normalsize
  Patrik Ed\'en\footnote{patrik@thep.lu.se}, G\"osta Gustafson\footnote{gosta@thep.lu.se}\\
  Department of Theoretical Physics\\
  Lund University\\
\end{center}
\vspace{5cm}
{\bf Abstract:} \\
We discuss some important issues concerning multiplicities in quark and gluon jets in $e^+e^-$ annihilation. In QCD the properties of a jet in general depends on two scales, the energy and virtuality of the jet. Frequently theoretical predictions apply to a situation where these scales coincide, while for experimental data they are often different. Thus an analysis to extract e.g.\ the asymptotic multiplicity ratio $\CF/C\srm A$ between quark and gluon jets, needs a carefully specified jet definition, together with a calculation of nonleading corrections to the multiplicity evolution. 

We propose methods to systematically study the separate dependence upon the two scales in experimental data and compare the results with theory. We present jet finding algorithms which correspond well to the theoretically considered jets. We also show that recoil effects add corrections to the modified leading log approximation which are quantitatively important, though formally suppressed at high energies.
\end{titlepage}

\section{Introduction}\label{sec:intro}
QCD predictions on the scale dependence of multiplicities in high energy \pair q systems are experimentally well confirmed~\cite{qqevol}. A similar multiplicity behaviour is to be expected from high energy gluon jets, the major difference being the different colour charges of gluons and quarks. 
In this paper we want to address a set of problems which have to be carefully treated for a quantitative analysis of jet properties:
\begin{itemize}
\item At experimentally accessible energies subleading effects are quantitatively important.
\item The jet properties depend in general on two scales, the energy of the jet and its virtuality, specified by the largest possible transverse momentum of one of its subjets. Theoretical calculations frequently refer to a situation where the two scales coincide, while in experimental analyses the two scales are often different.
\item In contrast to the topology of an event, which is mostly determined by a few energetic particles, the multiplicities in jets are sensitive to how the softer particles are associated to different jets.
\end{itemize}

As will be discussed in this paper, the multiplicity of hadrons depends not only  on the properties of the perturbative parton cascade, but also on the soft hadronization process. The assumption of Local Parton Hadron Duality (LPHD)~\cite{LPHD} implies a direct relation between the number of hadrons and the number of partons, provided a proper, locally invariant, cut-off is imposed on the parton cascade. The asymptotic behaviour of the multiplicity in jets was calculated in the leading log approximation in~\cite{LLA}. In the modified leading log approximation (MLLA), subleading terms in the evolution equations of relative magnitude $1/\sqrt{\ln s}$ are included~\cite{MLLA,pQCD}. With arguments based on ``preconfinement''~\cite{preconf} and LPHD we demonstrate in this paper that an important correction factor is expected due to recoil effects. Although formally suppressed by a factor $1/\ln s$, this effect is quantitatively large and has an essential impact on the ratio between quark and gluon jets at accessible energies.

Many jet finding algorithms have been presented for the study of $e^+e^-$ annihilation events~\cite{algs}. Several of these have been successfully used in comparisons between data and theory for properties like the distribution in the number of jets, and how this varies with the resolution scale. We want to stress that our problem is a different one, as described in the third point above, and it will be important to specifically consider the treatment of soft particles in the analysis.

The angular ordering effect in QCD~\cite{AO} implies that soft particles at large angles are emitted coherently from harder particles which they cannot resolve. Strictly speaking, these soft particles do not belong to any specific jet, and the colour factor for the emission is determined by the colour state of the combined unresolved partons. In ref~\cite{Cam} a method (the ``Cambridge'' algorithm) is proposed to associate the above-mentioned soft particles with the quark (or antiquark) jet, leaving to the gluon jet only those particles directly associated with the emitted gluon and the gluon colour charge. In this paper we study this question further and propose some modified cluster algorithms.

If we study two-jet events obtained in $e^+e^-$ annihilation using a jet finding algorithm with a distance measure of $k_\perp$-type, the jet properties (e.g.\ the hadron multiplicity) depend not only on the jet energy but also on the $k_\perp$-cut used. For minimum bias events, one hemisphere corresponds to a quark jet where these two scales are the same (an ``unbiased quark jet''). Similarly the multiplicity  of an unbiased gluon jet corresponds to one half of an imagined gg system stemming from a point source. In section~\ref{sec:jetscales} we will discuss how this quantity is related to a gluon jet in a \pair qg event. We will also present methods by which both of the jet scales can be systematically examined, as well as methods designed to define unbiased jets, where the two scales coincide.

Some experimental results of relevance for this paper have already been presented. In~\cite{OPALmethod,OPALres}, the hemisphere opposite to two quasi-collinear heavy quark jets in $e^+e^-$ events is analyzed. This corresponds well to an unbiased gluon jet, with $E\srm{g}\approx \sqrt s/2$. In~\cite{exps} the energy scale dependence of quark and gluon jets is studied. There  a fixed resolution scale is used in the cluster algorithm, which implies that the virtuality scale is held constant. In this paper, we discuss how to systematically study both scales in a jet. We also present methods to construct unbiased quark and gluon jets in a general three-jet topology, which enables a study of the scale evolution of unbiased jets.

In our analysis we will for convenience work in the Colour Dipole Model (CDM)~\cite{CDM}, which provides a geometric picture which is easily interpreted. The outline of this paper is as follows: In section~\ref{sec:LPHD} we discuss preconfinement and LPHD. In section~\ref{sec:CDM} we discuss the CDM and the multiplicity distributions, including corrections relevant to the MLLA approximation, as presented in~\cite{GG}. In section~\ref{sec:recoils} we discuss recoil corrections to the multiplicity evolution. In section~\ref{sec:jetscales} and \ref{sec:jetalgs} we discuss the two different scale dependences of multiplicities in jets and present cluster algorithms designed to extract these from data. In section~\ref{sec:results} we present results obtained by MC simulations. The results are discussed in section~\ref{sec:Summary}.

\section{Preconfinement and Local Parton Hadron Duality}\label{sec:LPHD}
It is essential to realize that the hadronic multiplicity cannot be determined from perturbative QCD alone. From perturbative QCD it is possible to calculate the parton multiplicity within a given phase space region. As the parton multiplicity diverges for collinear and soft emissions, some kind of cut-off is needed in order to find a quantity which is correlated to the hadronic multiplicity. The nature of this cut-off depends on the properties of the soft hadronization mechanism, and is therefore not calculable within perturbative theory.

The parton cascades are dominated by planar diagrams~\cite{planar} (to leading log level in an axial gauge). This implies that every colour charge after a cascade has a well defined partner anti-charge. As shown by Amati and Veneziano~\cite{preconf}, the final parton state can be subdivided in colour singlet clusters, whose masses stay limited also when the total energy becomes very large. In~\cite{cluster}, Marchesini, Trentadue and Veneziano showed that these ``preconfinement'' clusters are limited not only in momentum space but also in real space-time. At least in the large $\Nc$ limit it is then natural to assume that the clusters hadronize independently from the rest of the system. This means that the nonperturbative confinement mechanism is {\em local} in the sense that it combines partons which are directly colour connected and that it acts only locally also in momentum space.

This idea is further developed in the notion of Local Parton Hadron Duality (LPHD), proposed by the Leningrad group~\cite{LPHD}. Here a direct relation is assumed between partons and hadrons, which is {\em local} in phase space, which means that the dominant features of hadron distributions can be obtained from the parton cascades with an appropriate cut-off which is locally invariant.

\begin{figure}[tb]
  \hbox{
     \vbox{
	\mbox{
	\psfig{figure=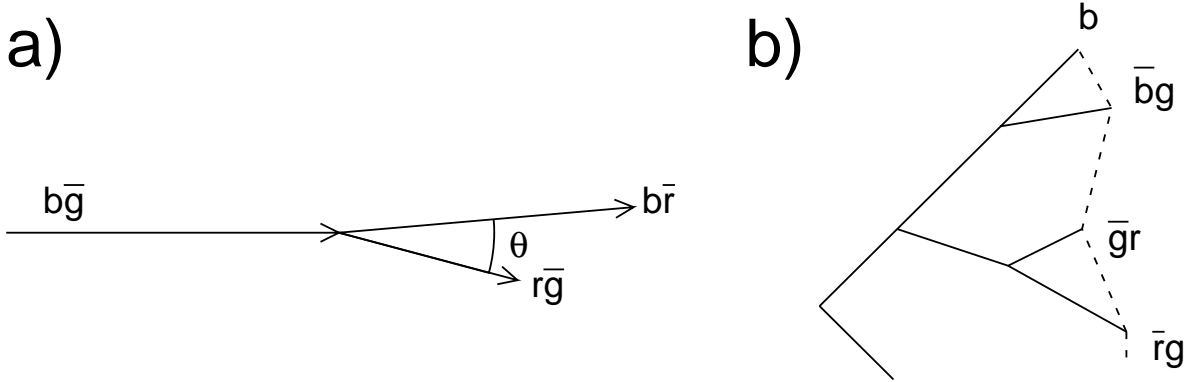,width=\figwidth}
	}
    }
  }
  \caption{{\bf a)} {\em After a g$\ra$gg emission with a small angle $\theta$, there is a screening of the new colour charges (\pair r in the picture). This implies that they give no net contribution to the emissions of soft gluons at polar angles larger that $\theta$ (the angular ordering effect). Their contribution to the emission at smaller angles corresponds to normal dipole emission in the \pair r back-to-back frame.} {\bf b)} {\em After a gluon cascade, further emissions of soft gluons corresponds to a set of independently emitting dipoles.}
}
  \label{f:smallangle}
\end{figure}
A very essential feature of the QCD cascade is ``soft gluon interference'' and ``angular ordering''~\cite{AO}. Study e.g.\ a b\anti g gluon emitting a r\anti g gluon thereby changing its own colour to b\anti r in a Lorentz frame where $\theta$ is small (see Fig~\ref{f:smallangle}a). In this case the red and antired charges screen each other in such a way that they do not give any emissions in directions with polar angles larger than $\theta$, the angle of the first emission. For these larger angles the two gluons emit softer gluons coherently as a single  b\anti g gluon, while for smaller angles the two gluons emit essentially independently. (For emission angles close to $\theta$ there is  some azimuthal asymmetry.) This implies that in a frame where the red and antired charges move back-to-back, their contribution to the emission of softer gluons is just the normal emission from a separating charge-anticharge pair.

The soft gluon interference (angular ordering) and a local cut-off are incorporated in the Marchesini-Webber formalism for the parton cascade~\cite{MWcasc}. This formalism is implemented in the HERWIG Monte Carlo~\cite{Web,herw}. In~\cite{Web} it is also shown that a locally invariant cut-off in the parton virtuality is essentially equivalent to a cut-off in transverse momentum relative to the emitting parent parton, if this is measured in a Lorentz frame where the parton energies are large and the angles small. The HERWIG MC also contains a cluster fragmentation model with local properties. In this model the  gluons  are at the cut-off level split into \pair q pairs, which are combined to colourless clusters, which finally decay into hadrons.

A cut-off in $k_\perp$ measured in the fast-moving (small angle) frame is approximately equivalent to a cut-off with the same  $k_{\perp \mrm{cut}}$ in the rest frame of the emitting charge-anticharge pair. In this ``dipole rest frame'' the emission is proportional  to $\frac{\mrm dk_\perp}{k_\perp}\mrm dy$, which  for a given $k_\perp$ means  a smooth rapidity distribution within the kinematically allowed region  
\eqbe \left|y\right|<\frac1 2\ln ( \hat s/k_\perp^2 ), \eqen
where  $\hat s$ is the squared mass of the dipole. This phase space corresponds in the fastmoving frame just to the region allowed by angular ordering. We see e.g.\ that for given $k_\perp$, the  rapidity range in the two jets is given by (for small angles $\theta$)
\eabe \Delta y_1+\Delta y_2 & = & \biggl[\ln (2p_1/k_\perp) - \ln(\cot \theta/2 )\biggr] +\biggl[\ln (2p_2/k_\perp) - \ln(\cot \theta/2 )\biggr] \approx \nonumber \\ & \approx & \ln(2p_1p_2(1-\cos \theta)/k_\perp^2) = \ln ( \hat s/k_\perp^2 ), \eaen
i.e.\ exactly the same result as in the dipole rest frame.
Thus in a gluonic cascade as in Fig~\ref{f:smallangle}b the emission of softer gluons corresponds to a set of independent ``dipole emissions''. A locally invariant cut-off is obtained by a  $k_{\perp \mrm{cut}}$ in the individual dipole rest frames. This is the basis of the Lund Colour Dipole Model~\cite{CDM}, which is discussed in more detail in the following section.

The local feature of the hadronization mechanism is also inherent in the Lund string fragmentation model~\cite{jet}. Here the hadrons which originate from the colour field stretched between a colour charge and its associated anticharge is independent of what happens further away in the system, with a correlation length corresponding to a few hadron masses~\cite{jim}. In ref~\cite{lambda} a measure, called $\lambda$, is proposed, which in the string fragmentation model is strongly correlated to the hadron multiplicity. With a local cut-off for the cascade, the $\lambda$-measure is also strongly correlated to the parton multiplicity~\cite{Dahl}. We note that in both cluster and string fragmentation there is a connection between a colour charge and its associated anticharge. Thus, although the parton and hadron distributions are strongly correlated, they are not exactly identical.

The colour coherence and the local properties of the hadronization process are fundamental features of the models implemented in the MC simulation programs HERWIG~\cite{herw}, ARIADNE~\cite{ARI} and JETSET/PYTHIA~\cite{JET}. The great phenomenological success for these programs in describing experimental data, in particular for $e^+e^-$-annihilation, is a strong support for the local features of the hadronization mechanisms expressed in preconfinement and LPHD. We note also that the independent jet fragmentation model, which does not have this feature, has not been able to describe the data in a satisfactory way.

In spite of the phenomenological success mentioned above, there are still fundamental open questions concerning the hadronization mechanism. In the large $\Nc$ limit there is a unique way to connect the partons as in Fig~\ref{f:smallangle}b. This is, however, not the case when $\Nc=3$. If two gluons have identical colours the confinement mechanism may connect partons which are not directly connected in the cascade generated by the simulation program. These ``colour reconnection'' effects are suppressed by $1/\Nc^2$, and some possible consequences are discussed in~\cite{recouple,CFexec,mixcouple}. No effects have, however, been observed so far in experimental data~\cite{OPALres}. Closely related to this problem is the possibility of colour reconnection between partons from the decay of different $W$:s in a $W^+W^-$ pair at LEP2~\cite{mixcouple,WWcouple}. This is of special interest as it might affect the $W$ mass determination, but also here no statistically significant effects have yet been found, e.g.\ in form of different decay multiplicity or modified Bose-Einstein correlations~\cite{WWdata}.

In perturbation theory it is possible to calculate the {\em parton} multiplicity within a given region of phase space. The conclusion of this section is that this is not the whole story if we want to calculate the multiplicity of {\em hadrons}. The hadronization effects of a parton containing e.g.\ a red colour charge depends upon where in phase space its partner antired charge is located, and this dependence is determined by the soft hadronization mechanism. An effective cut-off depends on the local properties of a jet and cannot be determined as a fixed phase space region valid for the whole jet. In~\cite{Mueller2} Gaffney and Mueller calculate the ratio of the parton multiplicity in quark and gluon jets within a fixed narrow cone, including correction terms of order $\alpha\srm s(Q^2)$ (or ${\cal O}(1/\ln Q^2)$). In section~\ref{sec:recoils} we show that, assuming a local hadronization mechanism based on preconfinement and LPHD, we expect a further correction term which, although suppressed by a factor $1/\ln Q^2$, is numerically important.

\section{The Colour Dipole Model}\label{sec:CDM}
\subsection{The Dipole Cascade}\label{sec:CDMcasc}
A high energy \pair q system radiates gluons with the distribution
\eqbe \mrm dn=C\srm F\frac{\alpha\srm s}{2\pi}\mrm dx_1\mrm dx_3\frac{x_1^2+x_3^2}{(1-x_1)(1-x_3)},\label{e:dnqq} \eqen
where $x_1$ and $x_3$ are the scaled quark and antiquark momenta, and $C\srm F = \frac1 2 \Nc(1-1/N^2\srm c)$.
With the definitions
\eabe 
\al=\frac6{11-2\Nf/\Nc},&~~~&\alq=\al\left(1-\frac1{\Nc^2}\right),\nonumber \\
k_\perp^2=s(1-x_1)(1-x_3),&~~~& y=\frac1 2\ln\left(\frac{1-x_3}{1-x_1}\right), \nonumber \\
\kappa= \ln(k_\perp^2/\Lambda^2),&~~~& \frac{\Nc\alpha\srm s(k_\perp^2)}{2\pi}=\frac\al\kappa,
\eaen
this distribution can be written
\eqbe \mrm d n=\frac{\alq}\kappa \mrm d\kappa\mrm dy\frac{x_1^2+x_3^2}2\approx \frac{\alq}\kappa \mrm d\kappa\mrm dy, \eqen
where the last approximation holds for soft gluons. 

\begin{figure}[tb]
  \hbox{
     \vbox{
	\mbox{
	\psfig{figure=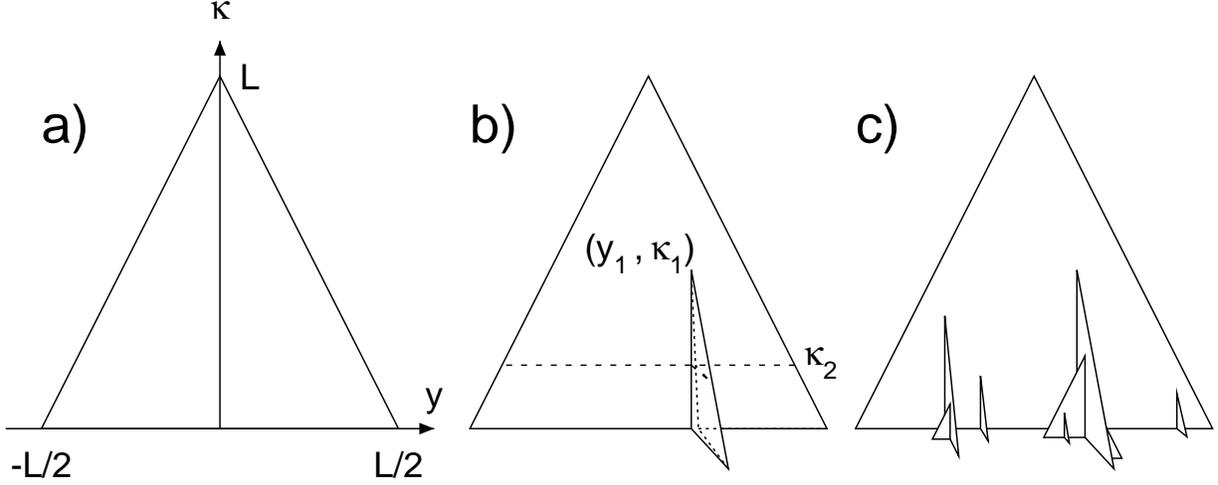,width=\figwidth}
	}
    }
  }
  \caption{{\bf a)} 
{\em The phase space for a gluon emitted from a \pair q dipole is a triangular region in the ($y$,$\kappa$)-plane ($\kappa=\ln~k_\perp^2/\Lambda^2$, $L=\ln~s/\Lambda^2$).} {\bf b)} {\em After one emission at $(y_1,\kappa_1)$, the phase space for a second (softer) gluon is represented by this folded surface.} {\bf c)} {\em Each emitted gluon increases the phase space for softer gluons. The total gluonic phase space corresponds to this multifaceted surface. The hadron multiplicity measure, $\lambda(L)$, is given by the length of the baseline.}
}
  \label{f:folds}
\end{figure}
The kinematical constraint $k_\perp<\sqrt s/(2\cosh y) \approx \sqrt s\exp(-|y|)$ implies that $\kappa+2|y|<L$, where $L$ is given by
\eqbe L=\ln(s/\Lambda^2). \eqen 
Thus the  allowed phase space for gluon emission is approximately a triangular region in the $\kappa,~y$-plane, cf. Fig~\ref{f:folds}a. 
After the emission of a gluon at $\kappa_1,~y_1$, the distribution for emissions of softer gluons corresponds to two independently emitting dipoles, one between the quark and the gluon, the other between the gluon and the antiquark. The available rapidity range for a gluon at $\kappa_2<\kappa_1$ is then $\ln(s\srm{qg}/k_{\perp2}^2)+\ln(s\srm{g\ol q}/k_{\perp2}^2) = L+\kappa_1-2\kappa_2$. Thus the phase space for further emissions can be represented by a folded surface as in Fig~\ref{f:folds}b. This can be generalized for several emissions and a multi-gluon event corresponds to a picture with many folds and sub-folds as in Fig~\ref{f:folds}c. 

\begin{figure}[tb]
\parbox{0.4\figwidth}{
  \hbox{ \vbox{
	\mbox{\psfig{figure=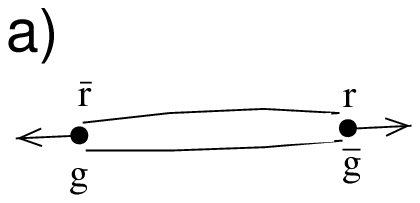,width=0.35\figwidth}}
   }  }
}
\parbox{0.5\figwidth}{
  \hbox{ \vbox{
	\mbox{\psfig{figure=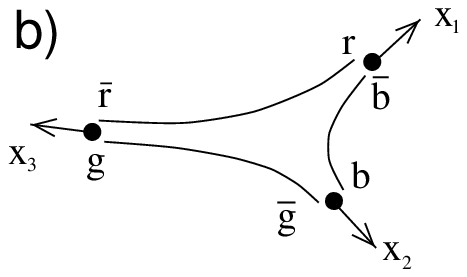,width=0.4\figwidth}}
   }  }
}
  \caption{{\bf a)} {\em An original two-gluon system corresponds to two colour dipoles.} {\bf b)} {\em After one emission, the emission of softer gluons (with lower $k_\perp$) corresponds to three dipoles.}
}
  \label{f:ggdipoles}
\end{figure}
In this language an initial gg system corresponds to two dipoles. If the gluons are e.g.\ r\anti g and g\anti r, we have a \pair r and a \pair g dipole as illustrated in Fig~\ref{f:ggdipoles}a. If one of these dipoles emits a gluon, the emission of softer gluons corresponds to three dipoles, as illustrated in Fig~\ref{f:ggdipoles}b, and after $n$ emissions we get a closed chain of $n+2$ dipoles. The phase space of an original gg system corresponds to  two triangular regions as in Fig~\ref{f:folds}a, glued together along the outer diagonal lines.

The emission from a dipole stretched between two gluons is however not exactly the same as from a dipole stretched between a quark and an antiquark. For a gg dipole the emission can in analogy to Eq~(\ref{e:dnqq}) be described by the distribution~\cite{CDM}
\eqbe \mrm dn=\frac{\Nc}2\frac{\alpha\srm s}{2\pi}\mrm dx_1\mrm dx_3\frac{x_1^3+x_3^3}{(1-x_1)(1-x_3)}. \label{e:dngg}\eqen
For soft and collinear emissions this goes over into the standard g$\ra$gg splitting function~\cite{CDM}. For soft emissions we have $x_1^3+x_3^3\approx 2$, which implies
\eqbe  \mrm d n\approx\frac{\al}\kappa \mrm d\kappa\mrm dy. \eqen
This result agrees with the soft emission from a \pair q dipole in Eq~(\ref{e:dnqq}) apart from the colour suppressed difference between $\alq$ and $\al$, i.e.\ between $\CF$ and $\Nc/2$.

\begin{figure}[tb]
  \hbox{  \vbox{
	\mbox{ \psfig{figure=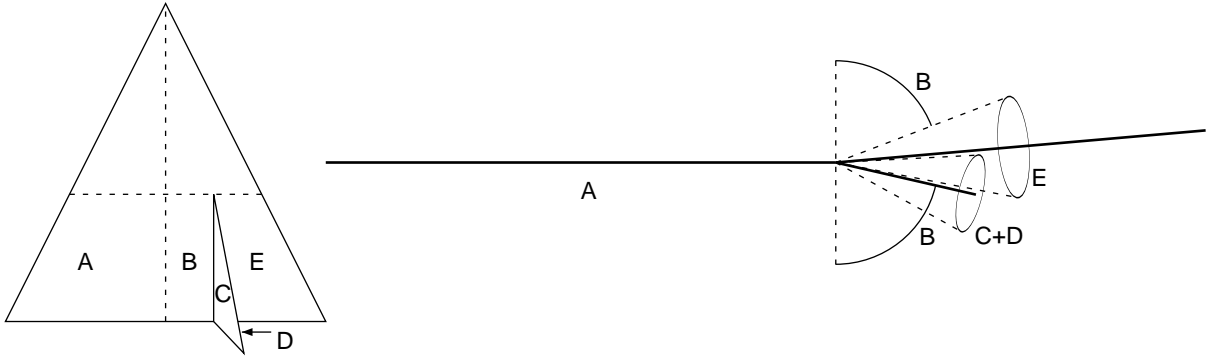,width=\figwidth}}
  } }
  \caption{
{\em After one gluon emission, the different regions in the folded $\kappa$-$y$ phase space approximately corresponds to the angular regions shown in the right figure. The angular directions in which the gluon and (anti)quark emits coherently approximately corresponds to region $B$, and the emission density there is proportional to $\CF$. Corrections to these region identifications are discussed in section~\ref{sec:OSC}.}
}
  \label{f:CFregions}
\end{figure}
After the first emission in a \pair q dipole, the different phase space regions for further emissions can be associated to different angular regions. Region $A$ of Fig~\ref{f:CFregions} roughly corresponds to emissions with negative rapidity in the overall CMS-frame, and region $B$ to particles with positive rapidity and a larger angle to the \pair q direction than the first gluon. Emissions from region $E$ have larger rapidity (smaller angle) than the first gluon. This is also the case for region $C+D$, with the rapidity measured in the gluon direction. The first emitted gluon and the (anti)quark will radiate coherently with the colour charge of the parent (anti)quark in region $B$. This argument can be generalized to a situation with several gluon emissions. For a cascade strongly ordered in $k_\perp$, this implies that the colour factor is $\CF$ in the original \pair q phase space triangle and $\Nc/2$ on all extra folds. The identification of regions presented in Fig~\ref{f:CFregions} is however only approximately true and we will in section~\ref{sec:OSC} study the corrections and their consequences in more detail. 

\subsection{Multiplicity Distributions}\label{sec:LLA}
\begin{figure}[tb]
  \hbox{ \vbox{
	\mbox{\psfig{figure=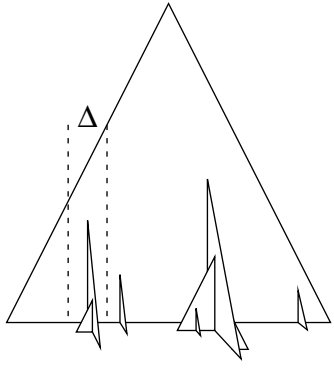,width=\figwidth}}
    } }
  \caption{\em The distribution $P$ can be subdivided into distributions for different rapidity intervals $\Delta$.}
  \label{f:PDelta}
\end{figure}
Assuming LPHD, the hadron multiplicity $N^h$ is closely related with the parton multiplicity $n^p$.
We will here briefly describe how the parton distribution $P(n=n^p,L=\ln(s))$ is derived in the dipole formulation~\cite{CDM}, in order to more easily discuss the effects of recoil corrections in the next section.
To find the  parton distribution $P(n,L)$, we first look at the distribution $P_\Delta(n)$ in a small rapidity interval $\Delta$, c.f.\ Fig~\ref{f:PDelta}.
The Laplace transform 
\eqbe \cP_\Delta(\gamma)\equiv \sum_n\exp(-\gamma n)P_\Delta(n) \eqen
has the property $ \cP_{\Delta_1+\Delta_2} = \cP_{\Delta_1}\cP_{\Delta_2}$. Thus
\eqbe \ln\cP(\gamma,L) = \sum_i\ln\cP_{\Delta_i}(\gamma). \label{e:sumlncP} \eqen
$P_\Delta(n)$ depends both on the width and the height of the interval $\Delta$. We denote the phase space height at rapidity $y$ by $l(|y|)$ and define
\eqbe R(\gamma,l)\equiv \lim_{\Delta\ra 0} \frac{\ln\cP_\Delta(\gamma)}\Delta. \label{e:Rdef} \eqen
Eq~(\ref{e:sumlncP}) can then be written
\eqbe \ln\cP(\gamma,L)=\int_{-y_\mrm{max}}^{y_\mrm{max}}\mrm dyR(\gamma,l(|y|)) = 2\int_{\kappa\srm c}^L\mrm dlR(\gamma,l)\left|\frac{\mrm dy}{\mrm dl}\right|. \label{e:dlnP0} \eqen
In~\cite{lambda}, it is shown that  $R$ also is related to $\cal P$ by
\eqbe \dL R^{(i)}(\gamma,L) = \frac{\alpha_i}{L}\left[\cP^{(\mrm g)}(\gamma,L)-1\right],~~~i=\mrm{q,g},~~~\alpha\srm g=\al. \label{e:dRc} \eqen
The boundary condition at some cut-off scale $\kappa_c$ for the cascade
\eqbe P^{(i)}_\Delta(n,\kappa_c) = \delta_{n0}~~ \Rightarrow~~R^{(i)}(\gamma,\kappa_c)=0, \eqen
then implies
\eqbe R^{(\mrm g)}(\gamma,L)=\frac{\al}{\alq}R^{(\mrm q)}(\gamma,L).\label{e:Rrat} \eqen
Combining Eq~(\ref{e:dlnP0}) and Eq~(\ref{e:dRc}) gives the following differential equation for $\cP$, valid in the LLA:
\eqbe \dLL\ln\cP^{(i)}(\gamma,L) = \frac{\alpha_i}{L}\left[\cP^{(\mrm g)}(\gamma,L)-1\right].  \label{e:dlnPcLLA}\eqen

\subsection{MLLA Corrections}\label{sec:MLLA}
\begin{figure}[tb]
  \hbox{ \vbox{
	\mbox{\psfig{figure=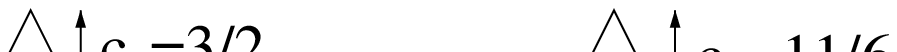,width=\figwidth}}
    } }
  \caption{
\em The hyperbolic shape of the true phase space limits and the inequality $x_1^3+x_3^3<x_1^2+x_3^2<2$ are both neglected in the leading order result. This can be corrected for by cutting off a strip at the triangle edges, thus reducing the available phase space. The different heights of the strips reflects the difference between $x_1^3+x_3^3$ and $x_1^2+x_3^2$, appearing in the emission density for a gg-- and \pair q dipole, respectively. The different magnitudes of phase space reductions implies $\nbar q(L)\sim\nbar g(L+c\srm g-c\srm q)$.
}
  \label{f:cgcq}
\end{figure}
Since the approximate triangular $(\kappa,y)$ region is somewhat larger than the true hyperbolic shape of the phase space, and the inequality $x_1^3+x_3^3<x_1^2+x_3^2<2$ was neglected, the emission density is overestimated in LLA (especially in a gg-dipole). In the Modified Leading Logarithmic approximation (MLLA)~\cite{MLLA}, corrections of relative order $1/\sqrt{L}$ are included. In~\cite{GG} it is shown, that evolution equations correct to relative order $1/\sqrt{L}$ are obtained if we maintain the approximation $x_1^3+x_3^3\approx x_1^2+x_3^2\approx 2$ and instead reduce the available phase space by cutting out a strip at the edges, as illustrated in Fig~\ref{f:cgcq}. The height of the strip is
\eqbe c\srm q = 3/2,~~~c\srm g = 11/6, \label{e:ci} \eqen
for a \pair q- and gg-dipole, respectively. The constant phase space reduction modifies $|\mrm dy/\mrm dl|$ to
\eqbe 2\left|\mrm dy/\mrm dl\right|^{(i)} = \Theta(L-c_i-l), \label{e:Theta} \eqen
which implies
\eabe 
\dL\ln\cP^{(i)}(\gamma,L) &=& \alpha_i R^{(i)}(\gamma,L-c_i), \label{e:dlnPcR} \\
\dLL\ln\cP^{(i)}(\gamma,L)& = &\frac{\alpha_i}{L-c_i}\left[\cP^{(\mrm g)}(\gamma,L-c_i)-1\right].  \label{e:dlnPc}\eaen
For the gluon case, the equations are modified by the possibility of g$\ra$\pair q splittings~\cite{GG}. We do not reproduce the algebraic details here, but refer to refs~\cite{MLLA,CDM}.
Extracting moments in $\gamma$ of Eq~(\ref{e:dlnPc}) leads to differential equations for  $\nbar g$ and $\nbar q$. Their asymptotic behaviours are
\eqbe \nbar q\approx (\alq/\al)\nbar g \sim L^\rho\exp(2\sqrt{\al L}), \label{e:nasympt}\eqen
where 
\eqbe \rho=\frac1 4-(2N\srm f/\Nc^3+11)\frac\al{12}\eqen
in MLLA. Thus the introduction of $1/\sqrt{L}$-suppressed terms in the evolution equations changes the asymptotic behaviour of $\nbar i$. 

The contribution to $\rho$ from the g$\ra$\pair q process is the term $2\Nf/\Nc^3$ in the parenthesis, which is clearly a small contribution compared to the other term 11. The contribution from this process to the ratio $N\srm q/N\srm g$ is also very small. Using numerical calculations we have found that it only modifies the ratio by less than 2\% for all energies above 4 GeV. For this reason we will neglect the splitting process in the analytical calculations presented below. However, when we compare our analytic results with MC simulations, the latter will include also the g$\ra$\pair q process.

Neglecting  the g$\ra$\pair q process for these reasons, we find from Eq~(\ref{e:dlnPc}) for the ratio  $N\srm q/N\srm g$ the MLLA result
\eqbe \frac{\al}{\alq}\nbar q(L) \approx \nbar g(L+c\srm g-c\srm q), \label{e:asympt1} \eqen 
which essentially is a $1/\sqrt{L}$ correction approaching the LLA result Eq~(\ref{e:nasympt}) for large $L$.

We also want to point out that the distribution in the multiplicity  measure $\lambda$, mentioned in section~\ref{sec:LLA}, satisfy exactly the same evolution equations as the distribution in $n$. Therefore the asymptotic increase is also the same. The boundary conditions at threshold are however different, which implies some deviations at lower energies.

\subsection{Virtuality Scale Dependence} \label{sec:virtscale}
\begin{figure}[tb]
  \hbox{ \vbox{
	\mbox{\psfig{figure=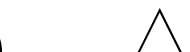,width=\figwidth}}
    } }
  \caption{\em {\bf a)} A two-scale dependent multiplicity can be studied in two-jet \pair q events, using a resolution $\kappa_r \ne L$.  The multiplicity is related to the phase space below the cut-off line at $\kappa_r$. The parton multiplicity is given by $\nbar i_{\alpha+\beta}=\nbar i(\kappa_r+c_i)$, and $\nbar i_\gamma = (L-\kappa_r-c_i)\frac{\mrm d}{\mrm d\kappa_r}\nbar i(\kappa_r+c_i)$. {\bf b)} If the hardest gluon jet is found at $\kappa_r$, the phase space which determines the multiplicity is given by the same region $\alpha+\beta+\gamma$, plus a fold. The multiplicity on this fold is given by $\nbar g(\kappa_r)$.}
  \label{f:trapets}
\end{figure}
If we consider a two-jet event sample, picked from \pair q events using some jet resolution scale $k_{\perp r}$, the mean multiplicity of the sample will be related to phase space area below $\kappa_r\equiv2\ln(k_{\perp r}/\Lambda)$.
As shown in Fig~\ref{f:trapets}, this implies
\eqbe \nbar q(L,\kappa < \kappa_r) \equiv \nbar q(\kappa_r+c\srm q)+(L-\kappa_r-c\srm q)\left.\frac{\mrm d}{\mrm dl}\nbar q(l+c\srm q)\right|_{l=\kappa_r} . \label{e:twoscale} \eqen 
Here the first term on the right hand side corresponds to the region $\alpha+\beta$ in Fig~\ref{f:trapets}a, while the second term corresponds to the region $\gamma$.
 Thus the mean hadron multiplicity, $N^h$, in \pair q events where no jet is resolved above $\kappa_r$ is given by
\eqbe \Nqq(L,\kappa < \kappa_r) = \Nqq(\kappa_r+c\srm q) + (L-\kappa_r-c\srm q)\frac{\mrm d}{{\mrm d}\kappa_r}\Nqq(\kappa_r+c\srm q). \label{e:NbelowK} \eqen
Events where the hardest gluon jet is found at  $\kappa_r$ corresponds to the phase space below the horizontal line in Fig~\ref{f:trapets}b. The hard gluon jet gives the contribution $\frac1 2\Ngg(\kappa_r)$ and therefore
\eqbe \Nqq(L,\kappa = \kappa_r) = \Nqq(L,\kappa < \kappa_r) + \frac1 2\Ngg(\kappa_r). \label{e:NatK} \eqen
We note that while the $\kappa_r$ dependence is fairly complicated, the $L$ dependence is simply linear.

\section{Recoil Effects and Boundary Conditions} \label{sec:recoils}
The ratio of the {\em parton} multiplicities in quark and gluon jets within a narrow cone was calculated including corrections to order $\alpha_s$, i.e.\ of order $1/L$, in ref~\cite{Mueller2} (c.f.\ also ref~\cite{Dremin}). The result is that the ${\cal O}(1/L)$ corrections are very small, only a few percent for energies above 30 Gev.

In a parton cascade, $\Lambda$ cannot be associated to a specific renormalization scheme, but must be treated as a free parameter. Changing $\Lambda$ introduces terms of relative suppression $1/L$ in the evolution equations Eq~(\ref{e:dlnPc}). Within the language of parton cascades, it is therefore non-trivial to systematically find a complete set of $1/L$ suppressed correction terms to many observables. 
We note however that the $1/L$ terms induced by changing $\Lambda$ are similar for the multiplicity in quark and gluon jets. Therefore these terms cancel to a large extent for the ratio $N\srm q/N\srm g$. This is in agreement with the result in ref~\cite{Mueller2}, that the correction term of order $\alpha_s$ to the ratio $N\srm q/N\srm g$ is small and scheme independent.

As discussed in detail in section~\ref{sec:LPHD}, the observable hadrons are not directly determined by the partons within a fixed phase space region. The hadrons originating from a particular parton colour charge depend also on the location in phase space of its corresponding anticharge. This implies that the necessary cut-off in the parton cascade should depend upon the location in phase space of an associated colour anticharge. Such a dependence appears naturally in the dipole cascade formalism with a $k_\perp$-cut in the dipole restframe. Therefore the result can be sensitive to recoil effects experienced by the emitting parent partons. Assuming the local properties of the hadronization mechanism discussed in section 2, we will in the next subsection derive a correction term to the ratio $N\srm q/N\srm g$. Although formally of order $1/L$, this recoil effect is numerically large, and important for comparisons between theory and experimental results.

In section~\ref{sec:results}, we compare our results with data from OPAL~\cite{OPALres} for jets with energy $M_Z/2$. In section~\ref{sec:jetalgs} we also discuss how to analyze LEP data at the $Z$ pole in order to obtain the ratio $N\srm q/N\srm g$ for a range of jet energies.

\subsection{Recoil Effects}
As discussed in section~\ref{sec:MLLA} we will neglect the effect caused by the g$\ra$\pair q process. Numerical calculations have shown that this process, which is both colour suppressed and kinematically suppressed, only has a negligible influence on the result. In principle the splitting process could be included at the expense of more complicated expressions, which would make the result less transparent.

Let us study the emission from an original gg-system. As discussed in section~\ref{sec:CDMcasc} this emission corresponds to two dipoles (a \pair r and a \pair g dipole in Fig~\ref{f:ggdipoles}a), and the phase space can be represented by two parallel triangular regions. If one gluon is emitted (e.g.\ from the \pair g dipole), the ensuing 3g state emits further soft gluons as three dipoles, as indicated in  Fig~\ref{f:ggdipoles}b. As discussed in section~\ref{sec:LPHD}, the emission from the separation between the b charge in the emitted gluon (2) and the \anti b charge in the emitting (recoiling) gluon (1) is determined by the invariant mass $s_{12}$. The Landau-Pomeranchuk formation time for an emitted quantum is determined by the $k_\perp$ of the emission. Thus there is a time ordering, such that a softer gluon (a gluon with smaller $k_\perp$) is emitted after the establishment of the configuration in Fig~\ref{f:ggdipoles}b. We must then expect that the weight of the \pair r and \pair g dipoles in  Fig~\ref{f:ggdipoles}b are determined by the corresponding dipole masses. The squared mass of the ``spectator'' \pair r dipole is reduced from its initial value $s$ to $s_{13}=(1-x_2)s$, which thus reduces the phase space for emissions from this dipole. We note that this recoil effect is very different for a \pair q system, where the weight of the corresponding dipole is negative, and suppressed by a factor $1/\Nc^2$. Therefore recoils will affect the ratio between quark and gluon jets, and we will now estimate the size of this effect.

The density of gluons with squared transverse momentum $k_\perp^2=s(1-x_1)(1-x_3)$ emitted by e.g.\ the \pair g dipole in Fig~\ref{f:ggdipoles}a is given by Eq~(\ref{e:dngg}).
This initial emission reduces the emission of softer gluons with $k^{\prime 2}_\perp<k_\perp^2$ in the partner \pair r dipole, so that the rapidity range is reduced from $\Delta y = \ln s/k^{\prime 2}_\perp$ to $\Delta y = \ln s_{13}/k^{\prime 2}_\perp = \ln s/k^{\prime 2}_\perp + \ln(1-x_2)$. On average the reduction $\delta y$ caused by gluons within an interval $\mrm d\ln k_\perp^2 \equiv \mrm d\kappa$ is then
\eabe \delta y &=& -\frac{\al}{\kappa}\mrm d\kappa\int\mrm dx_1\mrm dx_3\frac{x_1^3+x_3^3}{2(1-x_1)(1-x_3)}\ln(1-x_2)\delta(\kappa -\ln[s(1-x_1)(1-x_3)])\equiv \nonumber\\
&\equiv&\frac{\al}\kappa\mrm d\kappa\cdot I(L-\kappa),~~~(L=\ln s). \label{e:dy}\eaen
For a strongly ordered cascade we have $L\gg\kappa$, and in this limit we find
\eqbe I=2\left(\frac{\pi^2}6-\frac{49}{72}\right)\equiv c\srm r,~~~L-\kappa\gg1.\label{e:crdef}\eqen
Deviations from this limiting value give corrections which are suppressed at high energies. We have however also checked that the value in Eq~(\ref{e:crdef}) is a very good approximation in the whole relevant part of phase space. Typical values of $L-\kappa$ become smaller for smaller $L$--values, i.e.\ smaller dipole energies, due to the increase of $\alpha_s$. Within the MLLA approximation we find that for $L\sim6$ the average value of $\kappa$ becomes $\approx 2$, i.e. $k_\perp\approx e\Lambda$, which is very close to the standard cutoff in the ARIADNE cascade MC. Thus typical values of $L-\kappa$ are always larger than $4$, and for these values $I$ deviates from the value in Eq~(\ref{e:crdef}) by less than $10\%$.

Half of the effect in Eq~(\ref{e:dy}) corresponds to positive rapidities, in which case the recoil is taken essentially by the gluon (1) in Fig~\ref{f:ggdipoles}b, while in the other half the recoil is taken by gluon (3). The emissions in the \pair r dipole in Fig~\ref{f:ggdipoles}a has however a similar effect on the \pair g system, and the net result on one of the gluon jets is thus given by Eq~(\ref{e:dy}). 

Once one gluon is emitted two smaller dipoles are formed (between the pairs 3-2 and 2-1 in Fig~\ref{f:ggdipoles}b) emitting softer gluons, which give similar reductions of the ``spectator dipole'' phase space. Further down the emission cascade the original \pair g dipole in Fig~\ref{f:ggdipoles}a corresponds to a chain of dipoles. Emissions in the two end links of this chain give recoil effects on the spectator system originating from the \pair r dipole in Fig~\ref{f:ggdipoles}a. Consequently we obtain a recoil effect described by Eq~(\ref{e:dy}) for {\em all} values of $\kappa$, not only those which correspond to the first emission. Summing up the effects from all emissions above a given value $l=\ln k_\perp^2/\Lambda^2$, and replacing $I$ with its limiting value $c\srm r$, gives the total phase space reduction
\eqbe
\delta y\srm{tot}(l)=\int_l^{L-c\srm g}\mrm d\kappa\frac{\al}\kappa c\srm r = \al c\srm r \ln\frac{L-c\srm g}l. \label{e:dytot} \eqen
The upper limit is given by the kinematical limit in the MLLA approximation, $L-c\srm g$. Thus the effective rapidity range becomes
\eqbe \Delta y\srm{eff}(l)=L-l-c\srm g- \al c\srm r \ln\frac{L-c\srm g}l, \label{e:dyeff} \eqen
where $c\srm g$ is the MLLA correction and the last term is the recoil effect. We note in particular that the recoil term goes to zero at the kinematical limit $l=L-c\srm g$, as it should. From Eq~(\ref{e:dyeff}) we find
\eqbe \left|\frac{\mrm d\Delta y\srm{eff}(l)}{\mrm dl}\right|=1- \frac{\al c\srm r}l, \label{e:dydl} \eqen
which should be inserted in into Eq~(\ref{e:dlnP0}) (with $2y$ replaced by $\Delta y\srm{eff}$). Doing so, and taking the MLLA modification in Eq~(\ref{e:Theta}) into account, we find instead of Eq~(\ref{e:dlnPcR})
\eqbe \dL\ln\cP^{(\mrm g)}(\gamma,L)=\left(1-\frac{\al c\srm r}{L-c\srm g}\right)R^{(\mrm g)}(\gamma,L-c\srm g). \label{e:dlnPgrecoil} \eqen
We note in particular that the result in Eq~(\ref{e:dydl}) is independent of $L$ and that the correction in Eq~(\ref{e:dytot}) goes to zero in the kinematical limit. These two features imply that the structure of the evolution equation is unchanged when going from Eq~(\ref{e:dlnPcR}) to Eq~(\ref{e:dlnPgrecoil}).

For an initial \pair q system the emission from a quark colour charge is lower than one half of a gluon charge, $\CF=\frac{\Nc}2(1-1/\Nc^2)$. For the radiation from a \pair qg system, this can be expressed as a {\em negative} contribution with relative weight $-1/\Nc^2$ from a dipole stretched between the quark and antiquark, corresponding to the \pair r dipole in Fig~\ref{f:ggdipoles}b for the purely gluonic case. If we assume that we can treat this negative dipole in the same way, we would get the following result for a quark jet
\eqbe \dL\ln\cP^{(\mrm q)}(\gamma,L)=\left(1+\frac{\al \crqg q}{L-c\srm q}\right)R^{(\mrm q)}(\gamma,L-c\srm q), \label{e:dlnPqrecoil} \eqen
\eqbe \crqg q \equiv -\frac{1}{\Nc^2}\int_0^{1} \mrm dx\frac{1+(1-x)^2} {x} \ln(1-x)= \frac{2}{\Nc^2}\left(\frac {\pi^2}{6}-\frac{5}{8}\right). \eqen
Combining Eqs~(\ref{e:dlnPgrecoil}),~(\ref{e:dlnPqrecoil}) and~(\ref{e:Rrat}) now leads to
\eabe
\dL \ln\cP^{(\mrm g)}(\gamma,L+c\srm g-c\srm q) & = &\frac{\al}{\alq}\frac{1-\al\crqg g/(L-c\srm q)}{1+\al\crqg q/(L-c\srm q)}\,\, \dL\ln\cP^{(\mrm q)}(\gamma, L) \approx \nonumber \\ & \approx & \frac{\al}{\alq}\left(1-\frac{\al c\srm r}{L}\right)\dL\ln\cP^{(\mrm q)}(\gamma, L),\label{e:dlnPrecoil}\eaen
where $c\srm r$ now is modified by approximately 10\%, to
\eqbe c\srm r\equiv \crqg g+\crqg q = \frac{10}{27}\pi^2-\frac3 2 \eqen

The negative dipole in the \pair qg system is caused by interference effects due to identical colour charges. As discussed in section 2, these interference effects are also connected to the possibility of ``colour reconnection'' in the hadronization process. As these problems are still not solved, we regard the recoil effect for the quark jet as uncertain. Since this contribution is only 10\% of the total effect this is however not a serious problem here.

Taking the first moment in the Laplace transform variable $\gamma$, we find the relation Eq~(\ref{e:dlnPrecoil}) between $\dL\nbar g$ and $\dL\nbar q$ as well. Remembering that $\nbar g$ refers to a gluon jet while $\nbar q$ refers to a \pair q system, the relations between multiplicities in two-parton systems should be given by
\eqbe \dL \Ngg(L+c\srm g-c\srm q)=\frac{\Nc}{\CF}\left(1-\frac{\al c\srm r}{L}\right)\dL \Nqq(L). \label{e:asymptN} \eqen

\subsection{Boundary Conditions for Hadron Multiplicities}
The relation Eq~(\ref{e:asymptN}) has to be supplemented by appropriate boundary conditions. Extrapolating Eq~(\ref{e:asymptN}) down to too small values of $L$ would imply that the hadron multiplicity in a \pair q system would be significantly larger than that in a gg system. At low values of $L$, the hadron multiplicity is largely determined by the hadronic phase space and thus by the total available energy. This implies that at some threshold value $L_0$, we expect to have the relation
\eqbe \Ngg(L_0)\approx \Nqq(L_0)\equiv N_0, \label{e:Nboundary} \eqen
Here $L_0$, though larger than $\kappa\srm c$, should correspond to an energy of only a few GeV. 

The precise value of $L_0$ strongly depends on non-perturbative QCD effects. At low energies, the fact that a \pair q string contains two quarks while a gg string does not, may influence the ratio $\Nqq/\Ngg$. Thus the value of $L_0$ is sensitive to details in the fragmentation of low energy \pair q and gg systems, while $N_0$ to a large extent depends on how the primarily produced hadrons decay. In principle $L_0$ ought to be determined by experimental data from charmonium and bottonium decays. In our analysis, we have instead determined $L_0$ and $N_0$ from Monte Carlo simulations of the Lund String Fragmentation model, using the JETSET 7.4 computer program~\cite{JET}. We then get $L_0\sim 5.7$, corresponding to an energy $E_0=\Lambda\exp(L_0/2)\sim 4$GeV for $\Lambda=0.22$GeV. Given $\Nqq$ and $L_0$ from MC simulations, $\Ngg$ can be derived by numerical integration of the right hand side of Eq~(\ref{e:asymptN}). The dependence on the threshold behaviour can however be avoided if one studies how the multiplicity varies with increasing energy. This possibility is further discussed in section~\ref{sec:results}.

\section{Scale Dependences in Jets} \label{sec:jetscales}
In the following discussion we will use the notation of~\cite{pQCD} and~\cite{GG}, i.e.\ $N^h\srm g$ and $N^h\srm q$ denote multiplicities in jets, while $\Ngg$ and $\Nqq$ refer to multiplicities in two-parton systems. We have already introduced $L$ and $\kappa$ as logarithmic energy-- and virtuality scales for two-parton systems. The multiplicity in a jet $j$ with energy $E_j$ will be studied as a function of the logarithmic jet energy scale
\eqbe L_j \equiv \ln\left((2E_j)^2/\Lambda^2\right) \label{e:Ljdef}. \eqen 
The jet energy is multiplied by a factor $2$ for two ``cosmetic'' reasons: Then the scales coincide for a gluon jet when the simple condition 
\eqbe L_j=\kappa \eqen 
is fulfilled. (I.e.\ the relevant energy scale of this gluon jet is $E\srm g=k_\perp/2$, which is the MLLA result presented in~\cite{pQCD}.)
Furthermore, the multiplicity in a one-scale dependent jet is simply given by 
\eqbe N^h\srm p (L_j)=\frac1 2N^h\srm{pp}(L_j),~~~\mathrm{p=q,g}. \eqen

\begin{figure}[tb]
  \hbox{ \vbox{
	\mbox{\psfig{figure=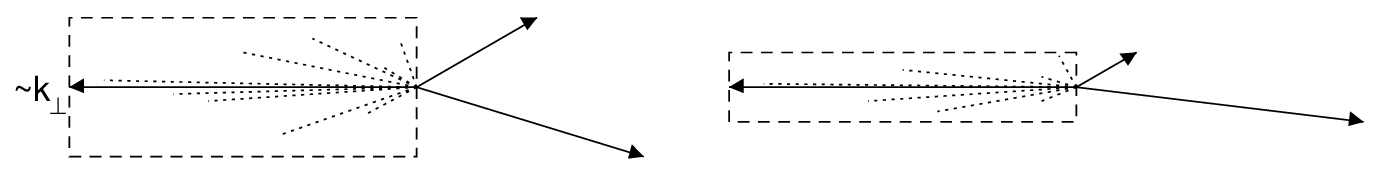,width=\figwidth}}
    } }
  \caption{\em The transverse resolution scale $k_\perp$ between the softer jets in a three-jet event also specifies the maximal allowed transverse momentum of unresolved particles within the jets. The multiplicity of the jets thus depends both on the jet energy and on $k_\perp$.}
  \label{f:twosc3jet}
\end{figure}
In general, the multiplicity of a jet depends on {\em two} scales, energy and virtuality. Reducing the maximal allowed transverse momentum within a jet reduces the multiplicity, even if the energy remains constant (c.f.\ Fig~\ref{f:twosc3jet}). 

As discussed in section~\ref{sec:virtscale}, the virtuality scale dependence of the multiplicity in quark jets can be easily studied in two-jet \pair q events varying the resolution scale. To compare quark and gluon jets we can use a three-jet event sample.  The multiplicity in the jets will also in this sample depend both on the jet energy and the resolution scale $k_{\perp r}$. Keeping $k_{\perp r}$ fixed, the energy scale dependence can be studied~\cite{exps}. However, the restriction implied by a fixed $k_{\perp r}$ is rarely considered in theoretical predictions. 
Rather, most calculations apply to the very forward region of jets~\cite{pQCD,Mueller2}. 
There coherence effects are negligible, and energy-momentum conservation restricts $k_\perp$ of emissions to such an extent that  $k_{\perp r}$ introduces no new bias. We will here examine the possibility to study unbiased jets at moderate energies, using the broadest possible cones to define the contents of the jets.

To find the jets of an event, cluster algorithms are used. In general, these contain a definition of  a distance measure $d$ between the jets, combine the two closest jets into one, and continues until all distances are above some resolution scale $\dcut$. In CDM and other approaches~\cite{herw}, the transverse momentum specifies the resolution. Thus it is appropriate to use a $k_\perp$-type of distance measure, as e.g.\ in the Durham~\cite{Dur}, LUCLUS~\cite{JET} or DICLUS~\cite{ARI} algorithms. We will use an approach, where jets are combined in the order given by the chosen algorithm until only three remain. The topology of the jets then specifies the transverse momentum, $k_\perp$, of the event (c.f.\ Fig~\ref{f:twosc3jet}). This approach enables the definition of unbiased jets, whose evolution with $k_\perp$ can be studied, while the energy scale dependence can be examined in an event sample with fixed $k_\perp$.

To construct unbiased jets, we note that a cone-like surface used to define the forward contents of a jet corresponds to a perpendicular plane in some other frame. In this frame the jet resembles one hemisphere of a full event, and if the energy and virtuality scales in this frame are equal, the jet is unbiased. We will refer to this observation as the ``One-Scale Criterion'' (OSC), since the mean multiplicity of the jet in this case indeed only depends on one scale -- the energy of the corresponding full event.

\section{Jet Algorithms}\label{sec:jetalgs}
Soft gluon coherence can be approximated by an angular ordering (AO) constraint~\cite{AO}. The multiplicity in a jet defined by AO will thus depend on one colour factor only. 
According to LLA, the jet will also be unbiased and depend on one single transverse energy scale, {\em provided} no extra cut-off in transverse momentum is imposed by a fixed jet resolution scale. We will present results from two algorithms based on this AO observation: The Cambridge algorithm presented in~\cite{Cam}, and a previously undiscussed ``Mercedes'' algorithm, where the multiplicities of the jets are defined in the symmetric ``Mercedes'' Lorentz frame of the event. The main reason for studying both is to look for similarities, which point at general properties in the AO approximation.

Corrections to the AO algorithms can be found by using the OSC explicitly. We will here discuss one such  algorithm, which has similarities with the Cambridge algorithm, called the ``Cone Exclusion'' algorithm, and one corrected ``Boost'' algorithm where the jets are analyzed in a frame similar to the Mercedes one. The two OSC algorithms give the same gluon jets, but the definition of quark jets differ. With the Cone Exclusion algorithm all three jets are unbiased, while the Boost algorithm constructs quark jets with well-defined but different energy and virtuality scales. Thus the Boost algorithm is suitable for a study of the separate dependence of the two scales, while the Cone Exclusion algorithm is suitable for a comparison of quark jets vs.\ gluon jets.

\subsection{Angular Ordering Algorithms}
\subsubsection*{The Cambridge Algorithm}
The Cambridge algorithm is designed to construct gluon jets uncontaminated by coherently emitted particles. Thus the gluon jet properties depend only on the transverse momentum to the nearest harder jet, and on the gluon colour factor $\Nc$. More specifically, the  Durham $k_\perp$-distance 
\eqbe d_{ij}\equiv 2\frac{\min\{E_i^2,E_j^2\}}{E\srm{vis}^2}(1-\cos\theta_{ij}) \eqen
is used to resolve jets, but the particles and sub-jets are merged in inverse angular order (those closest in angle are combined first). Once a soft jet is resolved, it is ``frozen out'', i.e.\ it gets no extra multiplicity contribution. Thus the contents of the soft jet is confined to a cone given by the smallest angle to any harder jet. This implies that the treatment of a gluon jet depends on wether its energy is higher or smaller than the quark jet energy. The Cambridge algorithm is therefore suitable for  gluon jet analyses primarily when the hardest gluon jet is significantly softer than the quark jets.

In the Cambridge algorithm, $\dcut$ not only determines when the clustering stops, but also at what stage different jets are frozen out. Producing a fixed number of jets is therefore not a completely trivial task. If we put $\dcut$ artificially large and pursue the clustering down to three jets, no freezing will occur. This would then correspond to a strict angular ordered clustering, similar to a cone algorithm. Fixing the number of jets is therefore better achieved by changing $\dcut$ in every event to a value which produces three jets. This procedure faces two problems: There may not be any $\dcut$ giving three jets, and when there is a large range of $\dcut$ values giving three jets, the multiplicities in the jets could depend on the choice of $\dcut$. However, both of these situations occur very rarely (at percentage level), and reliable conclusions may therefore be drawn from events clustered by the Cambridge algorithm, varying $\dcut$ to fix the number of jets to three.

\subsubsection*{The Mercedes Algorithm}
In conventional  cluster algorithms the bisectors between jets roughly separate the contents of them. In the specific case of a completely symmetric three-jet event, commonly referred to as a Mercedes event, such a jet definition will satisfy angular ordering~\cite{pQCD}. The kinematical constraint implies a fixed scale, but the scale evolution can still be studied by boosting a general three-jet event to its Mercedes frame. Since particles are in general shuffled from one jet to another under a Lorentz transformation, the mass of a jet is not invariant. Instead the {\em direction} of a jet, corresponding to a parton in the cascade, approximately transforms as a light-like vector. Using a $k_\perp$-based algorithm to find the jets, the Mercedes algorithm constructs gluon jets similar to those in the Cambridge algorithm.

\subsection{OSC Algorithms}\label{sec:OSC}
\subsubsection*{The Cone Exclusion Algorithm}
The ``Cone Exclusion'' (CE) method combines $k_\perp$- and angular distances in a way that has similarities with the Cambridge algorithm.
After the construction of three jets using a $k_\perp$-distance, a cone-like region is defined around the gluon jet. Only particles assigned to the jet that lie inside the region are then contributing to the multiplicity.  Note that in spite of the name ``Cone Exclusion'', the method is using a $k_\perp$-based cluster scheme to find the jets. The ``Cones'' are used to assign soft particles to the correct jet. 

With this method it is simpler to fix the number of jets to three than in the Cambridge algorithm, and it provides a better treatment of hard gluon jets. It is however specifically designed for  studies of multiplicities in jets in three-jet events, and does not share the benefits of the Cambridge algorithm as compared to other $k_\perp$-algorithms in other respects. E.g., the Cambridge algorithm is designed to avoid the formation of ``junk jets'' (when soft particles from different jets which happen to be close in phase space are combined and may be resolved as a jet if the resolution scale $\dcut$ is small). Since we in this analysis always pursue the clustering until only three jets remain, these ``junk jets'' will in general be absorbed into ``proper jets'', and we are therefore not very sensitive to this problem addressed by the Cambridge algorithm.

\begin{figure}[tb]
  \hbox{ \vbox{
	\mbox{\psfig{figure=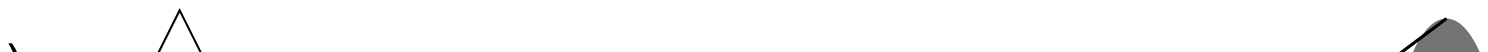,width=\figwidth}}
    } }
  \caption{{\bf a)} {\em If the partons of a dipole are going apart back-to-back, particles from the triangular region of height $\kappa_r$ in the direction of jet $j$ all have a rapidity $y>\ln(2E_j/\Lambda)-\kappa_r/2$, which implies $\cot(\theta_j/2)> 2E_j/k_{\perp r}$. This holds also if the dipole is not in its CMS.} {\bf b)} {\em If the dipole is in a general frame, the triangular phase space areas corresponds to two ``egg-shaped'' regions which after a boost along the bisector to the back-to-back frame become the cones of (a), i.e.\ which satisfy  $\cot(\theta^\prime_j/2)> 2E^\prime_j/k_{\perp r}$, where $\theta^\prime_j$ and $E^\prime_j$ are measured in the collinear frame.}}
  \label{f:cones}
\end{figure}
To find the proper cones to use, we will start our discussion with two-jets events obtained with some resolution $k_{\perp r}$. We look at the events in a frame where the original partons are going out back-to-back, though not necessarily in the overall CMS. In the $\kappa,y$ phase space picture (with folds), an unbiased jet  corresponds to a triangular region. When the partons of a dipole are moving apart back-to-back,  the relation $4E_1E_2=s$ implies
\eqbe L=\ln(\frac{2E_1}\Lambda)+\ln(\frac{2E_2}\Lambda) = \frac1 2(L_1+L_2), \label{e:jetscale} \eqen
where $L_1$ and $L_2$ are the logarithmic jet energy scales in this frame.
A particle p stemming from a triangular region of height $\kappa_r=2\ln(k_{\perp r}/\Lambda)$ in the direction of e.g.\ jet 2 has an angle $\theta\srm{p2}$ to the jet given by (c.f.\ Fig~\ref{f:cones}a)
\eqbe \frac1 2\ln\left(\cot^2\frac{\theta\srm{p2}}{2}\right)= y>\frac{L_2-\kappa_r}2=\ln\left(\frac{2E_2}{k_{\perp r}}\right). \label{e:2jetcones} \eqen 
An equivalent expression for this constraint is
\eqbe \ln(s\srm{p1})-\ln(s\srm{p2}) > \ln(s_{12})-\ln(k^2_{\perp r}). \label{e:invcones} \eqen

We now turn to three-jet events, where $k^2_\perp=s(1-x_1)(1-x_3)$ specifies the resolution. Then all quantities in the constraint in Eq~(\ref{e:invcones}) are Lorentz invariant, and it can immediately be applied to the three jets in the CMS. For the qg dipole, the constraints are then
\eqbe \frac{x\srm g\sin^2(\theta\srm{pg}/2)}{x\srm q\sin^2(\theta\srm{pq}/2)} < 1-x\srm q \label{e:CEg} \eqen
for the gluon jet and
\eqbe \frac{x\srm q\sin^2(\theta\srm{pq}/2)}{x\srm g\sin^2(\theta\srm{pg}/2)} < 1-x\srm q \label{e:CEq} \eqen
for the quark jet. Here $\theta\srm{pq}$ denote as before the angles between the hadron p and the jet direction q or g.
The relations for the \anti qg dipole are similar.

For the q (or \anti q) jet, Eq~(\ref{e:CEq}) specifies an ``egg-shaped'' region, which after a boost along the bisector of the dipole to a back-to-back frame becomes a cone satisfying Eq~(\ref{e:2jetcones}) (c.f.\ Fig~\ref{f:cones}). The gluon jet is however attached to two dipoles, and CE thus specifies two different regions. In the directions being accepted by one dipole but not the other, most particles emerge from the ``wrong'' dipole and should not contribute to the multiplicity of the gluon jet. We have required particles belonging to the gluon jet to satisfy both possible restrictions from Eq~(\ref{e:CEg}).

To summarize, the Cone Exclusion algorithm works as follows: Three jets are constructed, using a $k_\perp$-based cluster algorithm. A particle assigned to the gluon jet will then contribute to the multiplicity only if it satisfies Eq~(\ref{e:CEg}). The quark jets are treated similarly, using Eq~(\ref{e:CEq}). Thus the multiplicity in the forward region of every jet -- corresponding to an unbiased jet -- is studied, while soft central particles are simply ignored.

\subsubsection*{The Boost Algorithm}
The OSC can also be used to improve the AO-based Lorentz transformation to the Mercedes frame. Consider a three-jet event boosted to a Lorentz frame where the angles are
\eqbe \theta^\prime\srm{\ol qg}=\theta^\prime\srm{qg} \equiv \theta^\prime, \label{e:tprim}\eqen
and the partons carry energies $E_i^\prime$. Let the bisectors define the planes between different jet regions. For the gluon, the logarithmic back-to-back jet energy scale from both dipoles is then
\eqbe L\srm g=2\ln(\frac{2E^\prime\srm g}{\Lambda}\sin\frac{\theta^\prime}2), \eqen
which coincides with the virtuality scale when $L\srm g=\kappa$. This implies
\eqbe {x^\prime\srm g}^2\sin^2(\theta^\prime/2) = (1-x\srm q)(1-x\srm{\ol q}),~~~x^\prime_i\equiv\frac{2E^\prime_i}{\sqrt s}. \label{e:x2pr} \eqen
The search for cone-like boundaries which define an unbiased gluon jet has thus been reformulated to a search for a specific Lorentz frame where the jet regions are easily identified.

After a bit of algebra, one finds the general relation
\eqbe  {x^\prime\srm g}^2\sin^2(\theta^\prime/2) =(1-x\srm q)(1-x\srm{\ol q})\frac4{1-x\srm g}\cos^2(\theta^\prime/2). \eqen
Thus the requirement Eq~(\ref{e:x2pr}) is satisfied when
\eqbe \cos^2(\theta^\prime/2)=\frac{1-x\srm g}4. \label{e:cprim} \eqen
In the soft gluon limit, the wanted Lorentz frame coincides with the Mercedes frame, with $\theta^\prime=120^\circ$, but larger gluon energies give larger angles $\theta^\prime$. We also note that the energy scale for the quark jet in this frame is
\eqbe {x^\prime\srm q}^2\sin^2(\theta^\prime/2)=\frac{1-x\srm{\ol q}}{1-x\srm q}, \label{e:xipr} \eqen
and similarly for the $\mrm{\ol q}$ jet.

The {\em gluon} jet defined in this way is actually equivalent to the one defined by the CE algorithm. If we combine Eq~(\ref{e:x2pr}) and~(\ref{e:xipr})  into ${x^\prime\srm g}/{x^\prime\srm q}= 1-x\srm q$ and exploit the Lorentz invariance of the left hand side of Eq~(\ref{e:CEg}), we note that the CE condition for the gluon jet in Eq~(\ref{e:CEg}) can be rewritten as $\sin^2(\theta^\prime\srm{pg}/2)<\sin^2(\theta^\prime\srm{pq}/2)$, which is just the bisector condition in the Boost algorithm. The treatment of massive particles, whose masses have been neglected in the discussion, differ however in the two algorithms. Using both is a simple test of the sensitivity on particle masses.

The definition of quark jets differ in the two OSC algorithms.
From Eq~(\ref{e:xipr}), we see that the quark jets in the Boost algorithm  correspond to regions $A+B$ and $E$ in Fig~\ref{f:CFregions}. Thus the Boost algorithm provides means to study the two-scale dependent multiplicity for different jet energy scales in a fixed energy experiment, while the CE algorithm is better to use for the study of one-scale dependent multiplicities.

To summarize, the Boost algorithm is as follows: Find three jets using a $k_\perp$-based cluster algorithm. Boost the event to the frame where the jet directions, assumed to be massless, satisfies Eq~(\ref{e:tprim}) and Eq~(\ref{e:cprim}). Let the jet boundaries be given by the bisectors to the other jets in this new frame and re-assign particles to the jets accordingly.

\section{Results}\label{sec:results}
We have tested the analytic form for the ratio $\Nqq/\Ngg$ in Eq~(\ref{e:asymptN}) by comparing with multiplicity results from MC simulations and preliminary data~\cite{OPALres}. Using MC simulations, we also examine the presented jet algorithms by comparing the multiplicities obtained in jets with the multiplicity of complete events at corresponding energies.

The CDM is available as a Monte Carlo simulation program, ARIADNE~\cite{ARI}. There distribution factors such as $x_1^2+x_3^2$ are taken into account and energy conservation is obeyed. Thus Monte Carlo simulations can show whether corrections other than presented above are needed to understand the model predictions on the scale evolution. 

As discussed in section~\ref{sec:recoils}, a \pair qg system can be regarded as three dipoles, where the \pair q dipole is colour suppressed and has a negative weight. Alternatively, the system can be described by two dipoles, qg and g$\ol {\mrm q}$, where the colour factor transforms from $\CF$ in the q ($\ol {\mrm q}$) end to $\Nc/2$ in the gluon direction. However, dipoles with negative or non-uniform colour factors are ill suited for MC implementation. In the standard ARIADNE Monte Carlo, the solution is to neglect most terms of order $\Nc^{-2}$, using  the colour factor $N_c/2$ in all gg-- and qg dipoles. In~\cite{CFexec}, a modification to the MC correcting for this approximation is presented. There non-uniform colour factors in dipoles are implemented, in a way reflecting the discussion around Fig~\ref{f:CFregions}, where the emission density is assumed to be proportional to $\CF$ in all of the original \pair q phase space triangle, while $\Nc/2$ applies to all extra phase space folds after gluon emissions.

At moderate energies, the mean multiplicity is mostly determined by the hardest gluon. The emission of this gluon in the \pair q dipole is correctly given by $\CF$ in both MC versions. Thus, results from the two MC approaches are expected to deviate only at larger energies, showing a relative discrepancy of at most $1/\Nc^2$.

A ggg configuration is well described by three dipoles, all with positive emission density, determined by $\Nc/2$. This picture is implemented in default ARIADNE, which is used to obtain the simulation results from gg-systems presented below.

In the simulations we have used the default tune of ARIADNE 4.08, with $\Lambda=0.22$GeV. The parton configurations obtained in the cascade simulation are hadronized using the JETSET Monte Carlo~\cite{JET}, which is an implementation of the Lund String Fragmentation model.

\subsection{The Multiplicity Ratio $\Nqq/\Ngg$}  
From MC results we find the threshold energy where the multiplicity in quark and gluon jets are the same, to be given by $L_0\sim 5.7$. This corresponds to a CMS energy of $\sim 4$GeV. $L_0$ also specifies $N_0\equiv \Nqq(L_0)$.

\begin{figure}[tb]
  \hbox{ \vbox{
	\mbox{\psfig{figure=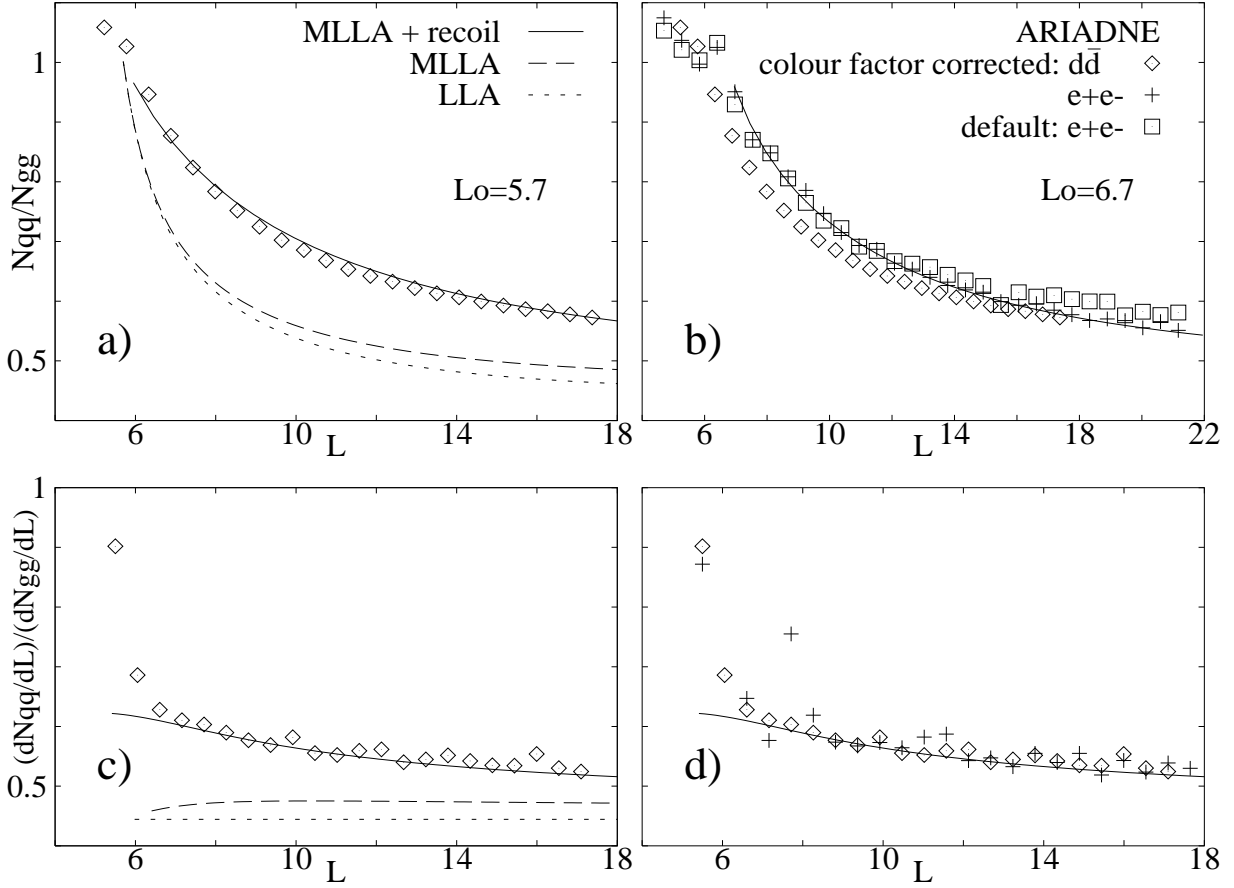,width=\figwidth}}
   }  }
  \caption{{\em Multiplicities ratios of \pair q and }gg {\em  systems, based on all stable particles. \pair q samples are based on pure \pair d events (diamonds) and $e^+e^-$ annihilation, using ARIADNE default (boxes) and the modification of~\protect\cite{CFexec}} {\em (crosses). The statistical errors for the MC results are within symbol sizes. We find that $\Nqq=\Ngg$ at $L_0=5.7$.} {\bf a)} {\em Comparison with the prediction of Eq~(\ref{e:asymptN}),} {\em using values of $N_0$, $L_0$ and $\Nqq(L)$ obtained from the MC. If neglecting the recoil correction (dashed line) and also the scale shift $c\srm g- c\srm q$ (dotted line), the prediction is far from MC. Thus recoil corrections of order $1/L$ are essential, even at $\sqrt s\sim 5$TeV.} {\bf b)} {\em For $e^+e^-$ events, \pair c and \pair b threshold effects are seen as a peak at $L\sim 6$ and a shoulder starting at $L\sim 8$. The relation determined by  Eq~(\ref{e:asymptN})} {\em fits well with $L_0=6.7$. The difference between default and modified MC are important at $L\sim 20$, but negligible at $Z_0$ energies and below.} {\bf c)} {\em The  $\dL N(L)$ ratio approaches the  asymptotic value $\CF/\Nc$ faster, but subleading corrections cannot be neglected.} {\bf d)} {\em The ratio of multiplicity derivatives $\dL N(L)$ is very similar for \pair d and $e^+e^-$ simulations and thus independent of $L_0$, as expected from Eq~(\ref{e:asymptN}).}}
  \label{f:ratios}
\end{figure}
In Fig~\ref{f:ratios}a, the simulated ratio of $\Nqq/\Ngg$ is compared to the prediction of Eq~(\ref{e:asymptN}), using values of $N_0$, $L_0$ and $\Nqq(L)$ obtained from the MC. If the $1/L$ recoil correction in Eq~(\ref{e:asymptN}) is neglected the result differs significantly, even at $L\sim 20$, i.e.\ at $\sqrt s\sim 5$TeV. The prediction is further modified if the MLLA correction of order $1/\sqrt L$ in Eq~(\ref{e:nasympt}) is neglected.

Fig~\ref{f:ratios}b shows the different multiplicity ratios obtained in simulations of $e^+e^-$ annihilation and pure \pair d events. At energies slightly above a heavy quark thresh-hold, the mean multiplicity gets a significant contribution from isotropically decaying heavy hadrons. This is seen as a peak at $L\sim 6$ ($\sqrt{s}\sim 2m_D$) and a little shoulder starting at $L\sim 8$ ($\sqrt{s}\sim 2m_B$). The details of these threshold effects are beyond the scope of this paper. Instead we note that Eq~(\ref{e:asymptN}) still fits well with $L_0\sim6.7$ ($\sqrt{s}\sim 6.3$GeV).
In Fig~\ref{f:ratios}b it is also shown that the corrections to $\Nqq$ from the Monte Carlo modification of~\cite{CFexec}, which more consequently implements the difference between $\CF$ and $\Nc/2$ after the first gluon emission, are important at very high $L$, but negligible  at $Z_0$ energies ($L=12$) and below.

The ratio between the derivatives $\dL N(L)$ is presented in Fig~\ref{f:ratios}c. This ratio is expected to approach the asymptotic value $\CF/\Nc$ more rapidly, which is also confirmed by data~\cite{exps}. This is also born out in our analysis, and we note a good agreement between MC and the analytic form in Eq~(\ref{e:asymptN}). According to this relation, the ratio of multiplicity derivatives $\dL N(L)$ is expected to be independent of $L_0$. Fig~\ref{f:ratios}d shows the similar results of \pair d and $e^+e^-$ simulations, confirming this expectation. 

In~\cite{OPALres}, the OPAL collaboration has studied the $\Ngg/\Nqq$ ratio via quark and gluon hemispheres with energies $E=M_Z/2$ ($L=12$). Their preliminary result
\eqbe \frac{\Ngg}{\Nqq}(L=12) =  1.509\pm0.022(stat)\pm0.046(syst) \eqen
is in excellent agreement with our result including recoil effects in Fig~\ref{f:ratios}b,
\eqbe  \frac{\Nqq}{\Ngg}(L=12) = 0.67 = \frac1 {1.5}. \eqen
In~\cite{OPALres}, OPAL also present preliminary data for the ratio of multiplicities in the central rapidity region,
\eqbe \frac{\Ngg}{\Nqq}(L=12,|y|<2) =  1.815\pm0.038(stat)\pm0.062(syst). \eqen
For a reasonably small central rapidity range $\Delta y$, the multiplicity $N(L,\Delta y)$ corresponds well to $\Delta y\dL N(L)$. It is therefore interesting to compare this result with the analytical expression for the ratio of multiplicity derivatives, which is more independent of the boundary conditions at $L_0$. From Fig~\ref{f:ratios}c, we find
\eqbe \frac{\dL \Nqq}{\dL \Ngg}(L=12) = 0.547 = \frac1 {1.83}, \eqen
with recoil effects included. The MLLA prediction without recoil effects is smaller than $1/2$.

Thus we conclude that MLLA calculations complemented with the recoil effect discussed in this paper are in very good agreement with the preliminary experimental data from OPAL. The recoil effect gives a sizeable correction which implies that the asymptotic value $\CF/\Nc$ is far beyond reach in accelerator experiments.

\subsection{Determination of Multiplicities in Jets}
The quantity $\Ngg$ in the previous subsection is the multiplicity in a gg event which is hard to realize and study directly in experiments. In section~\ref{sec:jetalgs}, different jet algorithms designed to investigate this quantity in normal three-jet events are presented. We have tested their performance by simulating events at $Z_0$ energies. The events are clustered into three jets and the multiplicities in the jets from different algorithms are studied as a function of $\kappa$ and $y$. The results for the jets are compared with the full-event results presented above. 

All generated events are considered in the MC analysis. Neutrinos are excluded from the event, all other neutral particles are treated as massless while all charged particles are treated as pions. The obtained visible system in each event is boosted to its CMS before the analysis. 

To tag the gluon jet, we note that the algorithms are applicable to a general jet topology. It is therefore possible to study events where one jet is much softer than the other two, when it is a good approximation to assume the softest jet to be the gluon jet. This makes the analysis simple and independent of sophisticated tagging methods. The gluon jet is softest in the phase space region $\kappa+2|y| < L-\ln(4)$. In our ``soft tag'' analysis, we have restricted the phase space to $\kappa+2|y| < L-2\ln(4)$, in order to avoid events with two similar soft jets. The restriction still allows us to study scales up to $k_\perp\sim 20$GeV.

In an experimental situation, harder gluon jets can be identified using heavy quark information, why it is of interest to test the performance of the presented  algorithms in a larger part of phase space. We have therefore also performed an analysis where the gluon jet is tagged using information available in the MC simulation, which however is experimentally non-observable. In this ``angle tag'' procedure, the jets are identified with the partons in such a way that the sum of jet-parton angles are minimized.

\subsubsection*{Multiplicities in Gluon Jets}
\begin{figure}[tb]
  \hbox{ \vbox{
	\mbox{\psfig{figure=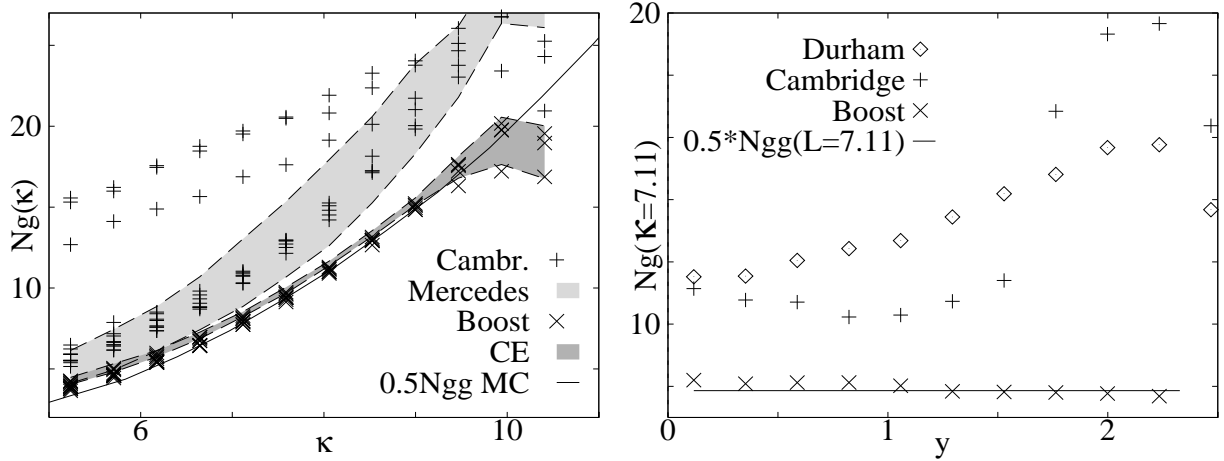,width=\figwidth}}
   }  }
  \caption{{\bf a)} {\em Multiplicities in gluon jets as a function of $\kappa$. Results for several values of $y$ are plotted, or the range of results are represented with shaded areas. The AO algorithms (Cambridge and Mercedes) have similar behaviours, and the multiplicity approaches $0.5\Ngg$ for low virtualities. The high values of the Cambridge algorithm corresponds to large $y$-values, where the gluon jet is not the softest one. The CE and Boost algorithms both give results independent of $y$ which are in very good agreement with $0.5\Ngg$ also for higher transverse momenta.} {\bf b)} {\em Multiplicities in gluon jets as a function of $y$ for fixed $\kappa$. The Durham algorithm assigns coherently emitted particles to the gluon jet, and the multiplicity increases with $y$. The Cambridge result is more independent of $y$, as long as the gluon jet is the softest one. $N^h\srm g$ of the Boost algorithms is very close to $0.5\Ngg$ for all $y$.}}
  \label{f:Ng}
\end{figure}
In Fig~\ref{f:Ng}a, the multiplicity of the gluon jet obtained by different algorithms is presented as a function of $\kappa$. For each $\kappa$, results for several values of $y$ are plotted. The gluon jet was identified with the ``angle tag'' method. The Angular Ordering algorithms (Cambridge and Mercedes) have similar behaviours. They perform well at low transverse momenta, but start to show a $y$-dependence at higher virtualities. This is especially so for the Mercedes algorithm. The OSC algorithms (CE and Boost) give results independent of $y$ and in very good agreement with $0.5\Ngg$.

In Fig~\ref{f:Ng}b, the multiplicity of the gluon jet for fixed $\kappa$ is presented as a function of gluon jet rapidity $y$. To allow the analysis to include large $y$, the ``angle tag'' method is used. With the Boost algorithm, $N^h\srm g(\kappa)$ is independent of $y$ and very close to the predicted $\frac1 2\Ngg(\kappa)$. The result from the Cambridge algorithm is independent of $y$ in a large range, but somewhat larger than $\frac1 2\Ngg$. The steep rise of the Cambridge multiplicity at $y\sim1.6$ reflects the different treatment of the gluon jet when it is not the softest one. A conventional cluster algorithm -- in this case the Durham algorithm -- assigns particles from the region of coherent emission to the gluon jet. For large $y$ this region becomes larger and the Durham multiplicity increases.

\begin{figure}[tb]
\parbox{0.47\figwidth}{
  \hbox{ \vbox{
	\mbox{\psfig{figure=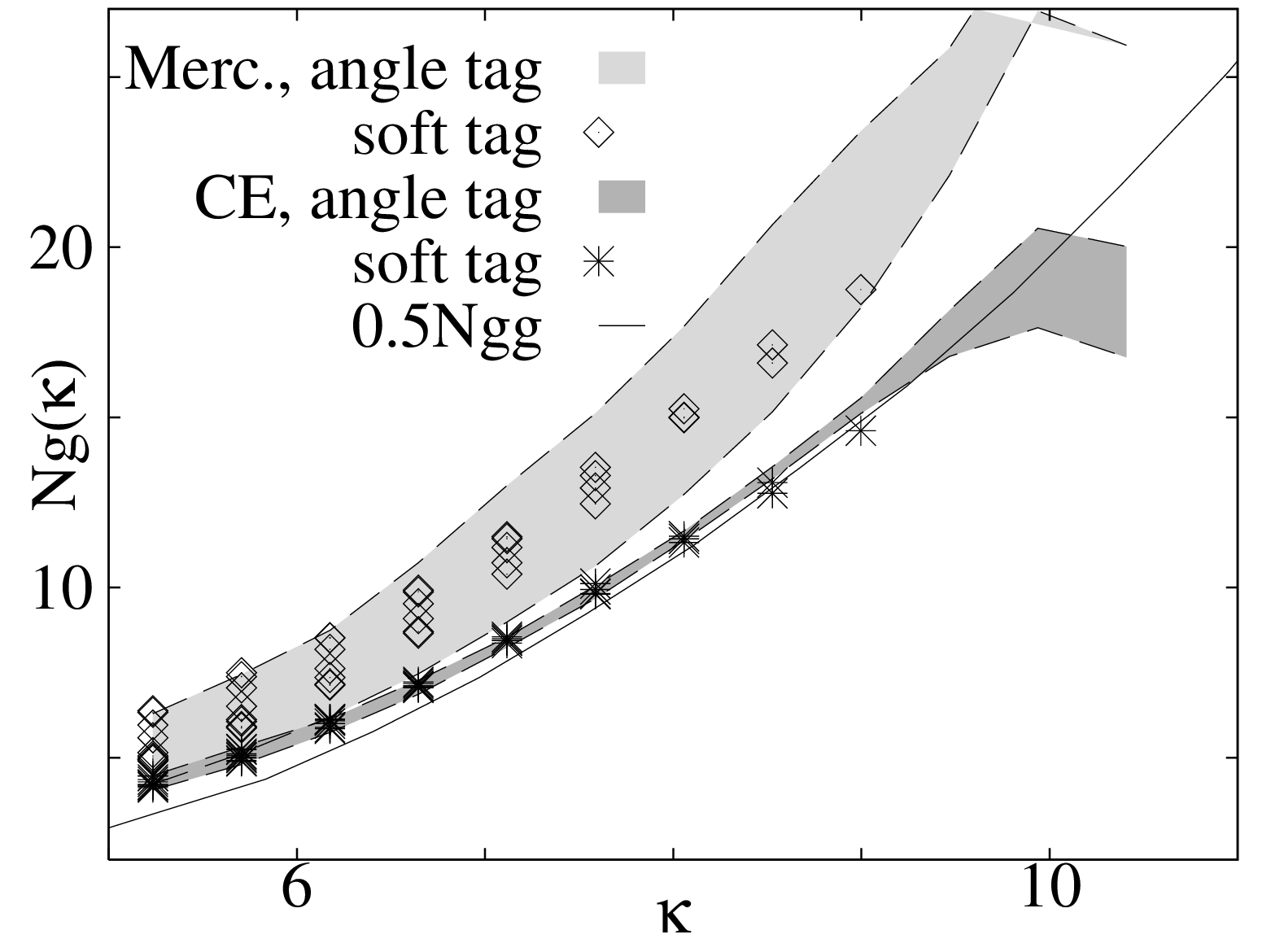,width=0.45\figwidth}}
   }  }
  \caption{\em Comparison of jet identification methods. The cluster algorithms presented are designed to properly construct gluon jets in a large part of phase space. Assuming the gluon jet to be softest (symbols) restricts the available phase space to some extent, but the result is in very good agreement with more sophisticated gluon tagging methods (shaded areas).}
  \label{f:Ngsoft}
}
\parbox{0.47\figwidth}{
  \hbox{ \vbox{
	\mbox{\psfig{figure=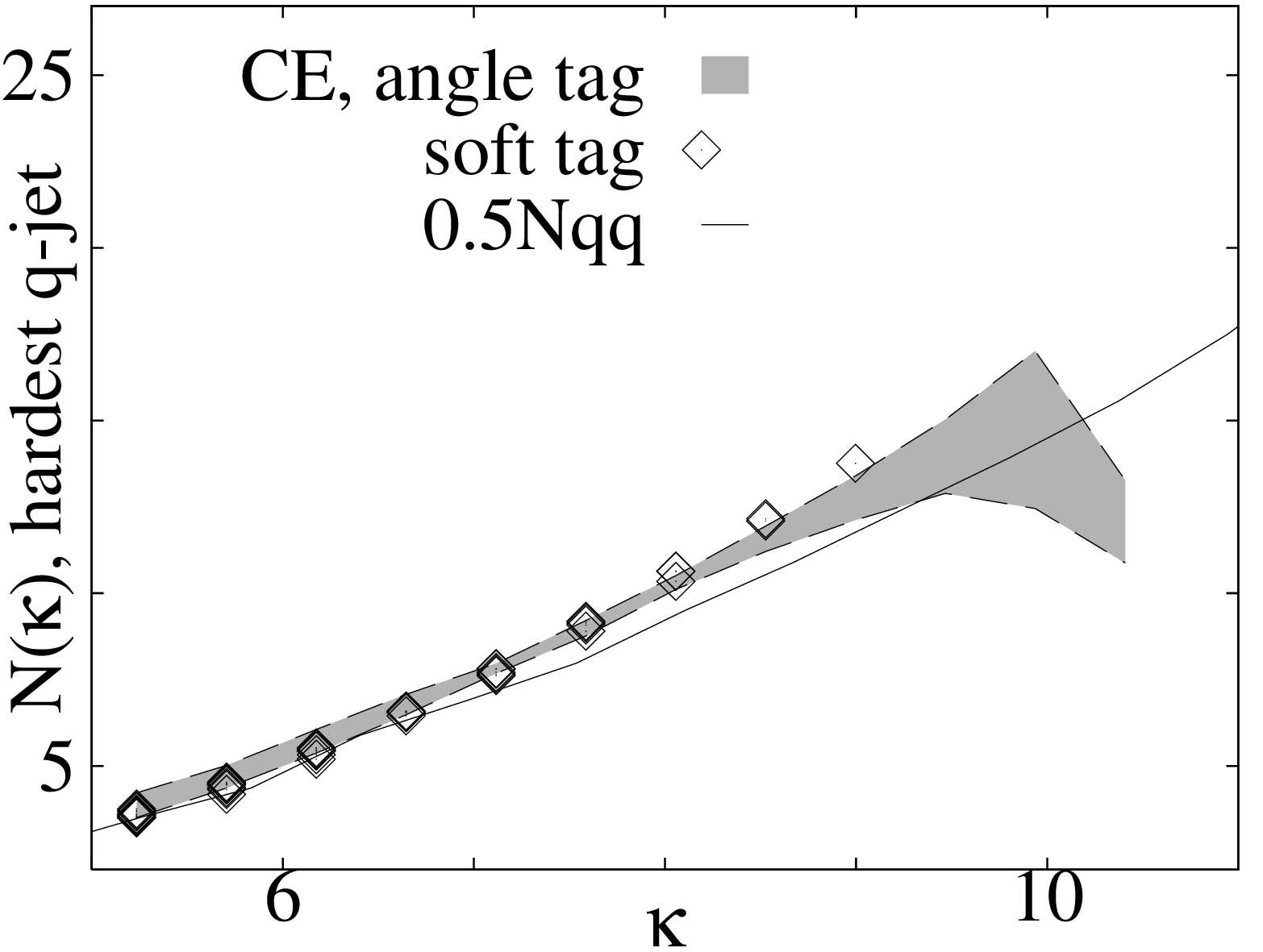,width=0.45\figwidth}}
   }  }
  \caption{\em Multiplicities in the hardest quark jet obtained by the Cone Exclusion method compared to half an $e^+e^-$ annihilation event. Results are plotted for several values of $y$. The range of results is given by a shaded region for the ``angle tag''. A reliable analysis can be performed in the kinematical region where the gluon jet is predominantly the softest one. }
  \label{f:Nq}}
\end{figure}
In Fig~\ref{f:Ngsoft} it is shown how the results using the very simple ``soft tag'' identification of gluon jets is in very good agreement with more sophisticated methods.

\subsubsection*{Multiplicities in Quark Jets}
In the Cone Exclusion algorithm, the OSC is used to define one-scale dependent regions also for the quark and antiquark jet. The results for the hardest quark jet are presented in Fig~\ref{f:Nq}. Again we note that jet mis-identifications in the ``soft tag'' method have little effect, and that this simple tagging procedure gives reliable results in a large kinematical region.

\subsection{Two-Scale Dependence}
\begin{figure}[tb]
	\mbox{\psfig{figure=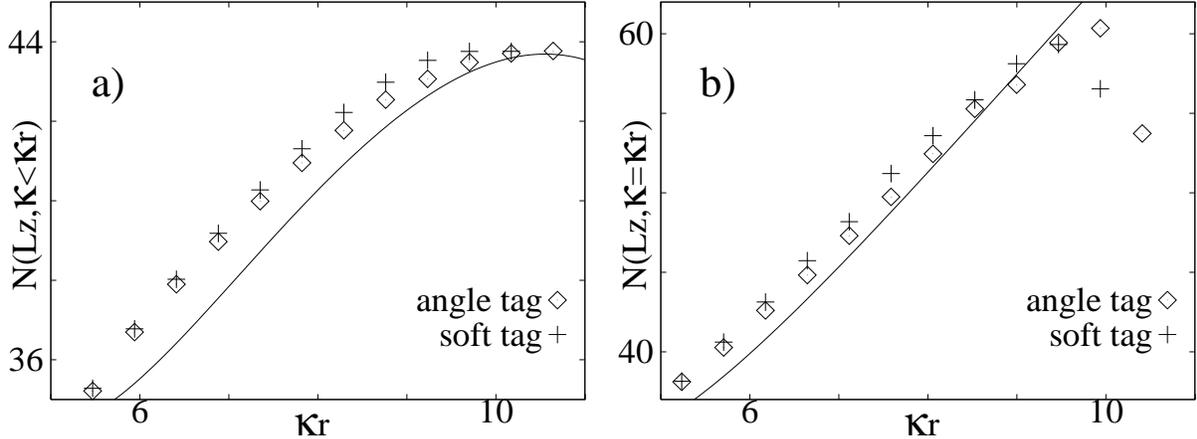,width=\figwidth}}
  \caption{
{\em  MC results of the multiplicity dependence on the jet resolution scale $\kappa_r$} {\em for $e^+e^-$ annihilation events with fixed CMS energy $\sqrt s=M_Z$. The gluon jet is identified both using the direction of the hardest gluon in the cascade and simply assuming the gluon jet to be softer than the quark jets. Solid lines are obtained using Eqs~(\ref{e:NbelowK})} {\em and~(\ref{e:NatK})}. {\bf a)} {\em The mean multiplicity in all events with no jet above $\kappa_r$} {\bf b)} {\em The mean multiplicity in all events with the hardest gluon jet at $\kappa_r$}.}
  \label{f:twoscale}
\end{figure}
The analysis required to compare data from fixed energy experiments with model predictions of the virtuality scale dependence is simple and straightforward. In Fig~\ref{f:twoscale} the mean multiplicity in events where no gluon jet is found using a resolution scale $\kappa_r$, and in events where the hardest gluon jet is found at $\kappa_r$ are presented. Since the full multiplicity of a two-jet or three-jet event is independent of how the particles are distributed among the jets, any $k_\perp$-based algorithm may be used. There is a fair agreement between MC simulations using the Durham algorithm, and the expectations from Eqs~(\ref{e:NbelowK}) and~(\ref{e:NatK}).

\begin{figure}[tb]
  \hbox{ \vbox{
	\mbox{\psfig{figure=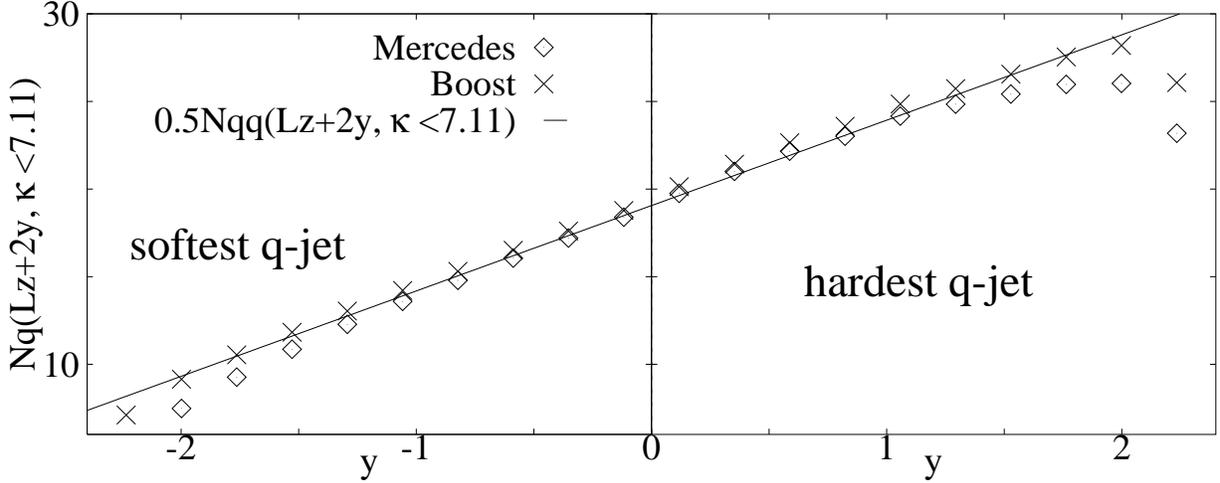,width=\figwidth}}
   }  }
  \caption{
{\em With the Mercedes and Boost algorithms, the energy scales of the quark jets are well defined and equals $\ln(s/\Lambda^2) \pm 2y$, where the rapidity $y$ of the gluon jet is defined by $y=\frac1 2\ln[(1-x\srm{\ol q})/(1-x\srm q)]$. Thus it is possible to study the energy dependence of the two-scale dependent multiplicity. MLLA predicts this dependence to be linear, which also is the result for the Boost algorithm, while the Mercedes algorithm (based on AO) deviates somewhat at large $|y|$.}}
  \label{f:Nqy}
\end{figure}
With the Mercedes and Boost algorithms, the multiplicities in quark jets are expected to be $0.5\Nqq(L+2y,\kappa<\kappa_r)$, where $y$ and $\kappa_r$ are the kinematical variables of the gluon jet. Thus the energy evolution of the two-scale dependence can be studied in a fixed-energy experiment. The expected linear dependence in $y$ (c.f.\ Eq~(\ref{e:NbelowK})) is seen in Fig~\ref{f:Nqy}, especially for the OSC-based Boost algorithm.

\section{Summary}\label{sec:Summary}

The ratio of the hadron multiplicity in quark and gluon jets is predicted to be 4/9 at very high energies. Corrections to this values calculated in the Modified Leading Log Approximation (MLLA) cannot expalin the difference between experimental data and this asymptotic value. In this paper we estimate the contribution from recoil effects to the ratio $N\srm q/N\srm g$. Although formally of order $1/\ln s$ this contribution is quantitatively sizeable. Combined with the MLLA expression the result agrees well with MC simulations and with preliminary experimental data for jet energies equal to $M_Z/2$~\cite{OPALmethod,OPALres}. We also discuss gluon jet definitions by which the scale evolution of gluon jets can be studied, using data from a fixed energy $e^+e^-$ experiment. These jet definitions thus should enable an extended analysis which would be interesting to combine with the results of~\cite{OPALres}. 

Another important point in this paper is that multiplicities in jets depend on {\em two} scales, the energy and the virtuality or maximal allowed transverse momentum. We derive expressions for the two-scale dependence and note that it can be examined with simple methods: Using a $k_\perp$-based cluster algorithm to construct exactly three jets, the multiplicities can be examined as a function of jet transverse momentum. 

We discuss the ``One-Scale Criterion'' (OSC), which states that a one-scale dependent jet will, in some Lorentz frame, correspond to one hemisphere of a two-parton event where the energy and transverse momentum scales coincide. The relevant scale for the jet is the energy of the corresponding hemisphere. Most of the commonly used cluster schemes of today concentrate on reproducing the jet energy and direction, and put less emphasis on the assignment of the soft particles to different jets. One exception is the Cambridge algorithm~\cite{Cam}, which based on Angular Ordering (AO) arguments constructs jets with one-scale dependent multiplicities. We present a set of algorithms designed to construct one-scale dependent jets. These are based either on AO or explicitly on the OSC. We examine the algorithms by analysing MC-generated events and comparing the obtained multiplicities in the jets with complete MC-simulated events at corresponding energies.

Our study shows that all the presented algorithms perform well, but that the OSC methods are somewhat better than the AO ones for quantitative analyses of multiplicities in jets. With the OSC algorithms the  gluon jet properties depend on only one scale. The treatment of quark jets differ, however. The ``Boost algorithm'' is particularly suited for studies of the two-scale dependence of quark jets, while the ``Cone Exclusion'' algorithm is designed for a study of one-scale jets, where the virtuality coincides with the energy of the jet.

We note that the algorithms can be used in a large kinematical region. We do not require the events to be Mercedes-like or Y-shaped. In events where one jet is significantly softer than the others, this jet is predominantly the gluon jet. Performing the analysis on these events makes it less important to identify the gluon jet via quark taggings.

\subsubsection*{Acknowledgments}
This work was supported in part by the EU Fourth Framework Programme `Training and Mobility of Researchers',
Network `Quantum Chromodynamics and the Deep Structure of Elementary Particles', contract FMRX-CT98-0194
(DG 12 - MIHT).

\end{document}